\documentclass[useAMS, usenatbib,fleqn]{mn2e} 
\usepackage{amsmath,environ}
\usepackage{times}
\usepackage{comment}
\usepackage{url}
\usepackage{amsmath}
\usepackage{graphicx}
\usepackage{subfig}
\usepackage{float}
\usepackage{caption}
\usepackage{amsfonts}
\usepackage{amssymb}
\usepackage{multirow}
\usepackage{color}
\usepackage[]{hyperref}
\usepackage{morefloats}
\usepackage{breqn}
\usepackage{color}

\pagerange{\pageref{firstpage}--\pageref{lastpage}} \pubyear{2013}
\def\LaTeX{L\kern-.36em\raise.3ex\hbox{a}\kern-.15em
    T\kern-.1667em\lower.7ex\hbox{E}\kern-.125emX}

\title[Large scale distribution of mass versus light from BAOs: search in the BOSS survey]%{Improved multi-parameter 
{Large scale distribution of mass versus light from Baryon Acoustic Oscillations: \\Measurement in the final SDSS-III BOSS Data Release 12}
\author[M. T. Soumagnac et al.]
{M. T. Soumagnac $^{1}$\thanks{E-mail: maayane.soumagnac@weizmann.ac.il}, C. G. Sabiu$^{2}$, R. Barkana$^{3}$, J. Yoo$^{4}$%\footnotemark[1]\thanks{This file has been amended to
\\
\\
$^{1}$Benoziyo Center for Astrophysics, Weizmann Institute of Science, 76100 Rehovot, Israel \\
$^{2}$Department  of  Astronomy,  Yonsei  University,  50  Yonsei-ro, Seoul 03722, Korea\\
$^{3}$Raymond and Beverly Sackler School of Physics and Astronomy, Tel Aviv University, Tel Aviv 69978, Israel\\
$^{3}$Astronomy Department, Harvard University, 60 Garden Street, Cambridge, MA 02138, USA\\
$^{4}$Korea Astronomy and Space Science Institute, 776, Daedeokdae-ro, Yuseong-gu, Daejeon, 34055, Korea\\
$^{4}$University of  Science and Technology (UST), Yuseong-gu 217 Gajeong-ro, Daejeon 34113, Korea
}

\begin{document}

\pagerange{\pageref{firstpage}--\pageref{lastpage}} \pubyear{2012}

\maketitle
\begin{abstract}
Baryon Acoustic Oscillations (BAOs) in the early Universe are predicted to leave an as yet undetected signature on the relative clustering of total mass versus luminous matter. This signature, a modulation of the relative large-scale clustering of baryons and dark matter, offers a new angle to compare the large scale distribution of light versus mass. A detection of this effect would provide an important confirmation of the standard cosmological paradigm and constrain alternatives to dark matter as well as non-standard fluctuations such as Compensated Isocurvature Perturbations (CIPs). The first attempt to measure this effect in the SDSS-III BOSS Data Release 10 CMASS sample remained inconclusive but allowed to develop a method, which we detail here and use to conduct the second observational search. When using the same model as in our previous study and including CIPs in the model, the DR12 data are consistent with a null-detection, a result in tension with the strong evidence previously measured with the DR10 data. This tension remains when we use a more realistic model taking into account our knowledge of the survey flux limit, as the data then privilege a zero effect. In the absence of  CIPs, we obtain a null detection consistent with both the absence of the effect and the amplitude predicted in previous theoretical studies. This shows the necessity of more accurate data in order to prove or disprove the theoretical predictions.
\end{abstract}

\label{firstpage}


\section{Introduction}

The imprint left by Baryon Acoustic Oscillations (BAOs), propagating in the baryon-radiation fluid before the time of recombination, is a powerful cosmological tool. The signature they left in the large scale distribution of mass was first detected in the 2dF Galaxy Redshift Survey (2dFGRS) \citep{Percival2001, Eisenstein2005, Cole2005} and was more recently measured in WiggleZ \citep{Blake2011} and the SDSS Data Release 12 \citep{Anderson2014}. 

Another important signature of BAOs is the imprint they left on the clustering of light {\it relative} to mass. Indeed, the acoustic waves which propagated in the baryon-radiation fluid before the time of recombination were only felt by the baryonic matter and not followed by Dark Matter (DM). After recombination, in the absence of any radiation pressure, gravitational instability took over the distribution of baryons and the strong discrepancy between the location of the baryonic shell and the cold-dark-matter started fading away. However, the resulting scale-dependency of the ratio of baryonic matter to total matter contrasts, $\delta_{\rm b}/\delta_{\rm tot}$, should still be observable at present time. Detecting this scale-dependency would offer a new angle to compare the large scale distribution of light versus mass, an effort which dates back to the 1980s \citep{Lahav1987, Erdogdu2006, Desjacques2016,Schmidt2016,Smith2017}. \cite{Soumagnac2016} conducted a first search for this effect in the data from the Baryon Oscillation Spectroscopic Survey (BOSS) data release DR10.

Specifically, the scale-dependency of $\delta_{\rm b}/\delta_{\rm tot}$, imprinted by BAOs, is important for three reasons:
\begin{enumerate}
\item The {\it detection} of the effect would provide a direct measurement of a difference in the large-scale clustering of mass and light and a confirmation of the standard cosmological paradigm. It would help rule out alternative theories of gravity, specifically non-DM models such as MOND \citep{Milgrom1994}. The main evidence against such theories today is the data from the {\it bullet cluster} \citep{Clowe2006}.
The measurement of the scale-dependency of  $\delta_{\rm b}/\delta_{\rm tot}$ from BAOs, would provide evidence comparable to the bullet cluster, with the additional advantage that this effect happens on linear scales.
\item The {\it amplitude} of the effect would allow a calibration of the dependence of the characteristic mass-to-light ratio of galaxies on the baryon mass fraction of their large scale environment.
\item \cite{Soumagnac2016} showed that such a detection would also allow constraints to be placed on the amplitude of Compensated Isocurvature Perturbations (CIPs).
\end{enumerate}

The measurement of the scale-dependency of $\delta_{\rm b}/\delta_{\rm tot}$ requires one to compare observable tracers of $\delta_{\rm tot}$ and $\delta_{\rm b}$. In this paper, we detail the approach by \cite{Soumagnac2016}, an extension of the proposal by \citet{Barkana2011}  (denoted BL11 in the rest of this paper) to use the number density $\delta_{\rm n}$ of galaxies as a tracer of the total matter density fluctuation $\delta_{\rm tot}$ and the absolute luminosity density of galaxies as a tracer of the baryonic density fluctuation $\delta_{\rm tot}$. In section~\ref{sec:mod}, we remind and detail the main aspects of the model developed by BL11 and extended in \cite{Soumagnac2016}. In section~\ref{sec:data}, we present our measurement of $\xi_{\rm L}$ and $\xi_{\rm n}$ from the SDSS-III BOSS Data Release 12 CMASS sample. Section~\ref{sec:fit} is dedicated to our model-fitting strategy. We then conclude on the significance of our detection with a model selection calculation, in section~\ref{sec:sel}. We give concluding remarks in section~\ref{sec:cclrem}. 
 
\section{The model}\label{sec:mod}

\subsection {A model for  $\delta_{\rm b}/\delta_{\rm tot}$}\label{sec:prediction}

The number density fluctuations $\delta_{\rm n}$ are driven by the underlying total matter density fluctuation $\delta_{\rm tot}$, with a bias $b_{\rm n}$, which should be approximately constant on large scales. 

\begin{dmath} \label{eq:deltan}
\delta_{\rm n}=b_{\rm n}\cdot \delta_{\rm tot}% + b_2(r(k)-r_{\rm lss})\cdot \delta_{\rm tot}\; ,%+\delta_{flux}
\end{dmath}

{\noindent On the other hand, an area with a higher baryonic mass fraction $\delta_{\rm b}/\delta_{\rm tot}$ than average is expected to produce more stars per unit total mass, hence more luminous matter, and to result in galaxies with lower mass-to-light ratio. As a result, the luminosity-weighted density fluctuation, $\delta_{\rm L}$, provides a tracer of $\delta_{\rm b}$, the baryonic contribution to $\delta_{\rm tot}$. }

Therefore scale-dependency of $\delta_{\rm b}/\delta_{\rm tot}$ induced by BAOs, should translate into a scale-dependency of $\delta_{\rm L}/\delta_{\rm tot}$. %where $\delta_{\rm L}$ is the absolute luminosity weighted density fluctuation
This being said, the mean luminosity of a given galaxy population relates to the baryonic content of the surrounding in a non-trivial way. The link between them is a combination of 
\begin {enumerate}
\item the way in which the luminosity of a galaxy depends on the baryon fraction of the host halo, 
\item the way in which the baryonic content of the host halo reflects the underlying baryonic contribution to the total matter density.
\end{enumerate}
%It depends on the baryon fraction in the halo, which in turn depends on the baryon fraction in the surroundings. But t

The luminosity density $\rho_{\rm L}$, for a given population of galaxies, is given by
\begin{equation} \label{eq:deltal}
\rho_{L}=n_{\rm gal}\left< L\right>
\end{equation}
where $\left<L\right>$ is the mean absolute luminosity of the population of galaxies.

$\left<L\right>$ may also depend on $\delta_{\rm tot}$, through the merger history of the population of galaxies. We model this dependency with a different bias $b_{\rm n}+b_{\rm L;t}$: 
\begin{dmath} \label{eq:deltal}
\delta_{\rm L}=(b_{\rm n}+b_{\rm L;t})\cdot \delta_{\rm tot}%+\delta_{flux}
\end{dmath}

This would be right if $\left<L\right>$ only depended on the large scale matter density. However, $\left<L\right>$ also depends on the baryon fraction in the host halo, $f_{\rm b}$. Following BL11, we assume that $\left<L\right>\propto(f_{\rm b})^{b_{\rm L;f}}$, where $b_{\rm L;f}\approx1.4$ is the bias factor of the luminosity density with respect to the halo baryon fraction.
Hence equation~\ref{eq:deltal} becomes 
\begin{dmath} \label{eq:deltal2}
\delta_{\rm L}=(b_{\rm n}+b_{\rm L;t})\cdot \delta_{\rm tot}+b_{\rm L;f}\delta_{\rm f}%+\delta_{flux}
\end{dmath}

\begin{figure}
\begin{center}
\includegraphics[width=6cm]{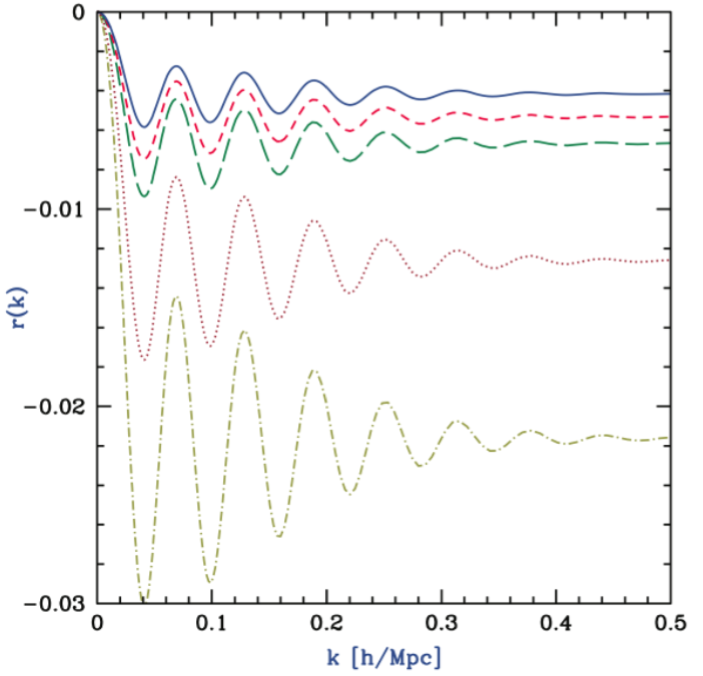}
\caption{The fractional baryon derivation $r(k)=(\delta_{\rm b}/\delta_{\rm tot})-1$, as a function of the scale $k$, at various redshifts ($z=0$,$0.5$,$1$,$3$ and $6$ from top to bottom). Credit: BL11.}
\label{fig:r_of_k}
\end{center}
\end{figure}

The link between the baryonic content of the halo  $\delta_{\rm f}$ and the baryonic content of the surrounding $\delta_{\rm b}$ is complex because of the non-linearity of halo collapse. It is derived in BL11 as,
\begin{dmath}
\delta_{\rm f}=\frac{A_{\rm r}}{\delta_{\rm c}}[r(k)-r_{\rm lss}]\delta_{\rm tot}\;,
\end{dmath}
\noindent
where
\begin{itemize}
\item $r(k)$ is the fractional baryon deviation $r(k)=\delta_{\rm b}/\delta_{\rm tot}-1$, shown in figure~\ref{fig:r_of_k} as a function of the scale k, and at various redshifts. $r(k)$ approaches a constant $r_{\rm lss}$ which depends on the redshift, on scales below the BAOs.
\item $\delta_{\rm c}$ is the critical total matter density $\delta_{\rm tot}$ of the halo at which the critical density of collapse is independent of mass and is equal to $1.69$ in the Eistein-De Siter limit, valid over a wide range of redshifts, \citep{Naoz2007}.
\item $A_{\rm r}$ is a corrective amplification factor coming from the use of the linear $r(k)$  in the non-linear halo collapse problem, and is expected to be $A_{\rm r}\approx3$, from simulations computed in BL11.

 \end{itemize}
Hence, the final equation for $\delta_{\rm L}$ is

\begin{dmath} \label{eq:deltal3}
\delta_{\rm L}=(b_{\rm n}+b_{\rm L;t})\cdot\delta_{\rm tot}+ b_{\rm L,\Delta}(r(k)-r_{\rm lss})\cdot \delta_{\rm tot}\;,
\end{dmath}
where  $b_{\rm L,\Delta}$ is a bias factor measuring the overall dependence of galaxy luminosity and the underlying difference between the baryon and total density fluctuations and is predicted in BL11 to be around $b_{\rm L,\Delta}\approx 2.5$.

%\subsubsection{Flux-limited surveys}\label{sec:fluxlimited}

In BL11, the authors show that, in the case of a flux limited survey, both equations \ref{eq:deltan} and \ref{eq:deltal3} must be slightly rethought. In such surveys, observed samples are limited by flux, or equivalently by luminosity if, for simplicity, we consider samples at a single redshift. 
The number of observable galaxies per unit of volume is given by
\begin{dmath}
F(L)=\int_{L'=L}^{\infty}\phi(L')dL'\;,
\end{dmath}
where $\phi$ is the luminosity function. The observed luminosity density of these galaxies becomes
\begin{equation}
\rho_{obs}=\left<L\right>F(L),
\end{equation}
where the mean luminosity of the sample $\left<L\right>$ is now defined as 
\begin{equation}
\left<L\right>=\frac{1}{F(L)}\int_{L'=L}^{\infty}L'\phi(L')dL'\;.
\end{equation}

One can then show that equations \ref{eq:deltan} and \ref{eq:deltal3} rewrite
\begin{dmath}\label{eq:finaldeltan}
\delta_{\rm n}=(b_{\rm n}+C_{\rm min}b_{\rm L;t})\delta_{\rm tot}+C_{\rm min}b_{\rm L,\Delta}[r(k)-r_{\rm lss}]\delta_{\rm tot}\;,
\end{dmath}
and
\begin{dmath}\label{eq:finaldeltal}
\delta_{\rm L}=\left(b_{\rm n}+(1+D_{\rm min})b_{\rm L;t}\right)\delta_{\rm tot} +b_{\rm L,\Delta}(1+D_{\rm min})[r(k)-r_{\rm lss}]\delta_{\rm tot}\;,
\end{dmath}

where $C_{\rm min}=\frac{L_{\rm min}\phi(L_{\rm min})}{F(L_{\rm min})}$ and $D_{\rm min}=\frac{L_{\rm min}}{\left<L\right>}C_{\rm min}$ with $\left<L\right>$ evaluated for $L=L_{\rm min}$.

The limit of a non flux-limitted survey corresponds to $C_{\rm min}=D_{\rm min}=0$. As noted in \cite{Barkana2011}, in the opposite limit where $L_{min}>L_{*}$ (L_{*} is the characteristic galaxy luminosity where the power-law form of the luminosity function cuts off), we can approximately set $\phi(L)\propto e^{-L/L_*}$ and then 
\begin{equation}\label{eq:CminLmin}
 C_{\rm min}=\frac{L_{\rm min}}{L_*}\;\;\;\;,\;\;\;\; D_{\rm min}=\frac{ C_{\rm min}L_{\rm min}}{L_{\rm min}+L_*}\,.
\end{equation}
\subsection{Compensated Isocurvature Perturbations} 

The measurement of the relation between dark matter and baryons, is related to the search for Compensated Isocurvature Perturbations (CIPs) \citep{Grin2014}. Measurements of primordial density perturbations are consistent with adiabatic initial conditions, for which the ratios of neutrino, photon, baryon and DM number densities are initially spatially constant. On the one hand, the simplest inflationary models predict adiabatic fluctuations \citep{Guth1982,Linde1982}. On the other, hand, more complex inflationary scenarii \citep{Brandenberger1994, Linde1984, Axenides1983} predict fluctuations on the relative number densities of different species, known as Isocurvature Perturbations (IPs). CMB temperature anisotropies limit the contribution of both baryons and DM to the total isocurvature perturbation amplitude. CIPs are perturbations in the baryons density $\delta_{\rm b}$ which are compensated for by corresponding fluctuations in the DM $\delta_{DM}$, so that the total density is unchanged. 

Such fluctuations are hard to detect, since the effects of gravity measurable by galaxy surveys (including galaxy numbers), only depend on the total density. Galaxy clusters gas fractions observations \citep{Holder2010} have lead to a weak constraints of the CIP's. $21$cm absorption observations are expected to allow a slightly better constraint of such perturbations \citep{Gordon2009}.
Under the standard assumption of a scale-invariant power spectrum for this field, equations~\ref{eq:finaldeltal} and~\ref{eq:finaldeltan} are modified to

\begin{dmath}\label{eq:finaldeltan2}
 \delta_{\rm n}=(b_{\rm n}+C_{\rm min}b_{\rm L;t})\delta_{\rm tot}+C_{\rm min}b_{\rm L,\Delta}[(r(k)-r_{\rm lss})\delta_{\rm tot}+\delta_{CIP}]\;,
\end{dmath}

 \begin{dmath}\label{eq:finaldeltal2}
 \delta_{\rm L}=(b_{\rm n}+(1+D_{\rm min})b_{\rm L;t})\delta_{\rm tot}+(1+D_{\rm min})b_{\rm L,\Delta}[(r(k)-r_{\rm lss})\delta_{\rm tot}+\delta_{CIP}]\;,
 \end{dmath}
where $\delta_{CIP}$ is a separate field that is uncorrelated with $\delta_{\rm tot}$.
With the method presented in this paper, we hope to improve the $10^{-2}$ current constraint on the amplitude of a scale invariant CIPs power spectrum \citep{Grin2014}.

\subsection{Model in terms of correlation function}%\label{sec:mod}

Equations~\ref{eq:finaldeltan2} and~\ref{eq:finaldeltal2} provide a model for the tracers of the quantities of interest $\delta_{\rm b}$ and $\delta_{\rm tot}$. However, the observable quantities in galaxy surveys are the two point statistics of such tracers, namely the power spectrum or the two-point correlation functions (2PCF). We reformulate the observational proposal of BL11 in terms of the 2PCF, defined as

\begin{dmath} \label{eq:xilin}
\xi({\bf x},{\bf y}) \equiv \int \frac{d^3{\bf k}}{(2\pi)^{3/2}} \frac{d^3{\bf k'}}{(2\pi)^{3/2}} \langle \delta( {\bf k} )\delta({\bf k'})\rangle e^{i{\bf k} \cdot {\bf x}} e^{i{\bf k} \cdot {\bf y}}=\frac{1}{2\pi ^2}\int k^2P(k)j_0(ks)dk \;,
\end{dmath}
where $P(k)$ is the power spectrum defined by $\left<\delta(k)\delta(k')\right>\equiv P(k)\delta^D(k-k')$.
In real space, and assuming $|r(k)-r_{\rm lss}|<<1$, equation~\ref{eq:finaldeltan} and equation~\ref{eq:finaldeltal} translate into the following,

\begin{eqnarray}\label{eq:xin_nosys}
  \xi_{\rm n} =B_{\rm n,t}^2\cdot\xi_{\rm tot}+ 
2B_{\rm n,t}B_{\rm n,\Delta}\cdot\xi_{\rm add}+
B_{\rm n,\Delta}^2B_{\rm CIP}\cdot\hat{\xi}_{\rm CIP}\;,\end{eqnarray}
and
\begin{eqnarray}\label{eq:xil_nosys}
\xi_{\rm L} =B_{\rm L,t}^2\cdot\xi_{\rm tot}+ 
2B_{\rm L,t}B_{\rm L,\Delta}\cdot\xi_{\rm add}+
B_{\rm L,\Delta}^2B_{\rm CIP}\cdot \hat{\xi}_{\rm CIP}\;.
\end{eqnarray}

 with 
 \begin{dmath}
 B_{\rm n,t}=b_{\rm n}+C_{\rm min}b_{\rm L;t}
 \end{dmath}
 \begin{dmath}
 B_{\rm n,\Delta}=C_{\rm min}b_{\rm L,\Delta}
  \end{dmath}
 \begin{dmath}
 B_{\rm L,t}=b_{\rm n}+(1+D_{\rm min})b_{\rm L;t}
  \end{dmath}
 \begin{dmath}  \label{eq:link}
B_{\rm L,\Delta}=(1+D_{\rm min})b_{\rm L,\Delta}=\frac{1+D_{\rm min}}{C_{\rm min}}B_{\rm n,\Delta}
  \end{dmath}
  
A key issue in the modeling of these correlation functions is to understand how they evolve with time. In section~\ref{sec:linear}, we adopt a simplistic approach and model $\xi_{\rm tot}$ and $\xi_{\rm add}$ with a linear perturbation theory. We then correct for the non-linearity of the clustering of galaxies in section~\ref{sec:nonlinear}.

\subsection{Linear-regime matter correlation}\label{sec:linear}
%\subsubsection{Power spectrum of $\delta_{\rm tot}$}

In order to model the $\xi_{\rm tot}$ and $\xi_{\rm add}$ components of equations~\ref{eq:xin_nosys} and~\ref{eq:xil_nosys}, we first compute a linear power spectrum $P(k)$ and a linear fractional baryon deviation $r(k)$ using CAMB \citep{Lewis2000}. We assume the same fiducial $\Lambda$CDM+GR, flat cosmological model with $\Omega_m= 0.274$, $h = 0.7$, $\Omega_bh^2 = 0.0224$, $n_s = 0.95$ and $\sigma_8 = 0.8$, matching that used by the BOSS collaboration in \citet{Anderson2014}. P(k) and r(k) are computed for the median redshifts of the sample we use, namely the CMASS sample of the BOSS DR12 release (\citealt{Ahn2012, Ahn2014,Alam2015}; see section~\ref{sec:data}). %two 
%samples we use, % WHY? Note that this model is different from the current best-fit cosmology; however these parameters allow us to translate\citet{Anderson2014}
%The computed $P(k)$ and $r(k)$ are shown in figure~\ref{fig:pkrk}.
%\begin{figure}
%\begin{center}
%\includegraphics[width=6cm]{./figs/baryonic_fraction_DR10.eps}
%\caption{$r(k)$ computed with CAMB \citep{Lewis2000}, shown at the DR12 median redshift ($z=0.570$).}
%\label{fig:pkrk}
%\end{center}
%\end{figure}

%\subsubsection{Power spectrum of $\delta_{CIP}$}
To model $\xi_{CIP}$, we make the standard assumption %TO DO REF
that the power spectrum of the CIP field is of the form $P_{CIP}(k)=A_{CIP}k^{-3}$ \citep{Grin2014}. Since the corresponding correlation function
 \begin{equation*}\label{eq:xicipnodamp}
 \xi_{CIP}(s)=\frac{1}{2\pi ^2}\int k^2P_{CIP}(k)\cdot j_0(ks)dk =\frac{A_{CIP}}{2\pi ^2}\int \frac{j_0(ks)}{k}dk
 \end{equation*}
diverges, we compute the integration from $k=10^{-4}h/Mpc$, which is also the minimum value of the CAMB linear power spectrum we use to model $\xi_{\rm tot}$.

\subsection{Corrections to the linear correlation functions}\label{sec:nonlinear}
  
The matter correlation $\xi_{\rm tot}$ predicted by linear perturbation theory does not exactly describe the clustering of galaxies. Nonlinear gravitational collapse and redshift distortions modify galaxy clustering relative to that of the linear-regime matter correlations, changing the shape of the correlation function. In particular, according to linear perturbation theory, the acoustic signature increases in amplitude but the characteristic scale imprinted in the early universe remains unaltered, whereas non linear growth of structure leads to a shift of the acoustic peak. 
In \cite{Soumagnac2016}, the authors accounted for two systematic effects due to nonlinear clustering: damping of the BAO peak %(section~\ref{sec:damping}) 
and mode coupling.% (section~\ref{sec:mode}).

\subsubsection{Damping}\label{sec:damping}

Simulations have shown that nonlinear structure formation and, to a lesser extent, redshift distortions erase the higher harmonics of the acoustic oscillations. This degrades the measurement of the acoustic scale. This effect is accounted for by ``damping''  the linear theoretical BAO on small scales. The damping term is often approximated by a Gaussian smoothing \citep{Percival2010}. The corrected correlation function is given by

\begin{dmath}
\xi_{\rm tot}(s)=\xi(s)\otimes e^{-(k_{*}\cdot s)^2}=\frac{1}{2\pi ^2}\int k^2P(k) e^{-(k_{*}\cdot k)^2}j_0(ks)dk \;.
\end{dmath}
The damping is also applied to $\xi_{\rm add}$ and $\xi_{CIP}$. %The effect of the damping on $\xi_{\rm tot}$, $\xi_{\rm add}$ and $\xi_{CIP}$ is shown in figure~\ref{fig:ingredients}.

\subsubsection{Mode coupling}\label{sec:mode}
Mode coupling generates additional oscillations that are out of phase with those in the linear spectrum, leading to shifts in the scales of oscillation nodes defined with respect to a smooth spectrum. When Fourier transformed, these out-of-phase oscillations induce percent-level shifts in the acoustic peak of the two-point correlation function. The corresponding correction to the damped linear correlation function is given in \citet{Crocce2008}, as:
\begin{dmath}
\xi_{\rm tot}(s)=\xi(s)\otimes e^{-(k_{*}\cdot s)^2}+A_{\rm MC}\xi'(s)\xi^{(1)}(s)\;,
\end{dmath}
where $\xi(s)$ denotes the linear correlation function of equation~\ref{eq:xilin} and 
\begin{dmath}
\xi^{(1)}(s)=\int\frac{d^3k}{k}P(k)j_1(ks)\;,
\end{dmath}
where $j_1$ is the first order Bessel function.

\subsubsection{Systematics}
The systematic effects of BOSS data are investigated in \citet{Ross2012} and essentially cause a constant shift in $\xi$.  A simple way to account for systematics that would affect  differently $\xi_{\rm L}$ and $\xi_{\rm n}$ is to add a constant to the model in equation~\ref{eq:xil_nosys}. Thus, equation~\ref{eq:xil_nosys} becomes
%changes that are quite close to constant shifts and there are theoretical systematics (e.g., the integral constraint) causing close to constant shifts.
\begin{dmath}
\xi_{\rm L} =B_{\rm L,t}^2\cdot\xi_{\rm tot}+ 
2B_{\rm L,t}B_{\rm L,\Delta}\cdot\xi_{\rm add}+
B_{\rm L,\Delta}^2B_{\rm CIP}\cdot \hat{\xi}_{\rm CIP}\\
\nonumber
+B_{\rm sys,L}\;.
 \end{dmath}

\subsubsection{Full model equations}
Our final model equations, also given in the supplemental material of \cite{Soumagnac2016} are:
\begin{eqnarray}\label{eq:xin_new}
  \xi_{\rm n} =B_{\rm n,t}^2\cdot\xi_{\rm tot}+ 
2B_{\rm n,t}B_{\rm n,\Delta}\cdot\xi_{\rm add}+
B_{\rm n,\Delta}^2B_{\rm CIP}\cdot\hat{\xi}_{\rm CIP}\\
\nonumber
+B_{\rm sys,n}\;,\end{eqnarray}
and
\begin{eqnarray}\label{eq:xil}
\xi_{\rm L} =B_{\rm L,t}^2\cdot\xi_{\rm tot}+ 
2B_{\rm L,t}B_{\rm L,\Delta}\cdot\xi_{\rm add}+
B_{\rm L,\Delta}^2B_{\rm CIP}\cdot \hat{\xi}_{\rm CIP}\\
\nonumber
+B_{\rm sys,L}\;.
\end{eqnarray}

where
\[
%\begin{eqnarray}
\xi_{\rm tot}(s)=\xi(s)\otimes e^{-(k_{*}\cdot s)^2}+
A_{\rm MC}\xi'(s)\xi^{(1)}(s)\;,%\]
%\end{eqnarray}
\]
\[
\xi_{\rm add}(s)=\frac{1}{2\pi ^2} \left(\int k^2[r(k)-
r_{\rm lss}]P(k) j_0(ks)dk \right)\otimes 
e^{-(k_{*}\cdot s)^2}\;,
\]
\[\xi_{\rm CIP}(s)\equiv B_{\rm CIP}\cdot \hat{\xi}_{\rm CIP}(s)=
 \frac{B_{\rm CIP}}{2\pi ^2}\left(\int \frac{j_0(ks)}{k}dk\right)\otimes
 e^{-(k_{*}\cdot s)^2}\; ,\] 
 %\[\hat{\xi}_{\rm CIP}(s) \equiv \frac{1}{B_{\rm CIP}}\xi_{\rm CIP}=
 %\frac{1}{2\pi ^2}\left(\int \frac{j_0(ks)}{k}dk\right)\otimes
 %e^{-(k_{*}\cdot s)^2}\; ,\] 
 where $\otimes$ denotes convolution, $\xi(s)$ is the linear
 correlation function (eq.~3 of the main text), and
\[
\xi^{(1)}(s) = \int \frac{d^3 k}{k} P(k) j_1(ks) \;. 
\]
Thus, our full set of parameters is $\theta=\{B_{\rm n,t},B_{\rm
  L,t},B_{\rm n,\Delta},B_{\rm L,\Delta},B_{\rm CIP},B_{\rm
  sys,n},B_{\rm sys,L},A_{\rm MC},k_*\}$.

%\begin{figure}
%\begin{center}
%\includegraphics[width=6cm]{./figs/Xitot.eps}
%\includegraphics[width=6cm]{./figs/Xiadd.eps}
%\includegraphics[width=6cm]{./figs/Xicip.eps}
%\caption{The three ingredients of the model, $\xi_{\rm tot}$,  $\xi_{\rm add}$ and  $\xi_{cip}$, shown at the DR12 CMASS sample median redshift ($z=0.57$).  All are shown for two different values of the damping parameter $k_*$,  presented in section~\ref{sec:damping}, and $\xi_{\rm tot}$ is also shown for different values of the mode coupling parameters $A_{\rm MC}$, presented in section~\ref{sec:mode}.}
%\label{fig:ingredients}
%\end{center}
%\end{figure}

In order to compute the oscillatory integral $\xi_{\rm tot}$, $\xi_{\rm add}$ and $\xi_{CIP}$, we wrote a Python wrapper for the fftlog code from \citet{Hamilton2000}.

\subsection{Previous results}\label{sec:previous_results}

In this section we summarize the results of the measurement by \cite{Soumagnac2016}, which used the DR10 data and are brought for comparision throughout the paper.

When allowing CIPS in the model, i.e., $B_{\rm CIP}\neq0$, the authors obtained evidence at $3.2\sigma$ of the relative clustering signature. The $1\sigma$ range of $1.1<B_{\rm L,\Delta}<2.8$ was consistent with the prediction of BL11 of $B_{\rm L,\Delta} \approx 2.6$ (predicted along with two assumptions: (1) $B_{\rm n,\Delta} \approx 0$ - an assumption that may be wrong here, as explained in section \ref{sec:constrained} - and (2) $B_{\rm n,t}$ and $B_{\rm L,t}$ approximately equal). In addition, the best-fit value of $B_{\rm CIP}$ is $2.3\times 10^{-3}$, with a $2\sigma$ upper limit of % 21 bins $B_{\rm CIP}=$B_{\rm CIP}=6.76\times 10^{-3} and 2 sigma limit is $3.18\times 10^{-2}$,
$B_{\rm CIP}=6.4\times 10^{-2}$, which is within an order of magnitude of the best existing limits noted previously. A full tabulation of the DR10 best-fit parameters is given in the Supplemental Material of \cite{Soumagnac2016}.
  
However, the DR10 results were not robust enough for making strong claims. When modeling the data without allowing for CIPs (i.e., setting $B_{\rm CIP} = 0$), the evidence for a detection of non-zero $B_{\rm L,\Delta}$ goes away. The authors obtained a null detection consistent with both the absence of the effect and the BL11 prediction.

\subsection{Luminosity function and constraint on $B_{\rm L,\Delta}$}\label{sec:constrained}

Within the model by BL11, the parameters and in particular the ratio $B_{\rm L,\Delta}/B_{\rm n,\Delta}$ depend on the flux limit of the survey. Here, we use equation~\ref{eq:link} and our knowledge of the BOSS sample to derive an additional constraint on the parameters of the model. The BOSS DR12 CMASS sample data are in the regime of rare, bright galaxies, well into the exponential tail of the luminosity function. 
Specifically, the flux limit in the $i$ band is $17.5 < i < 19.9$, which translates into $L_{\rm min}\approx8.4$\,$10^{10}L_{\rm \odot}$. With $L_*\approx2$\,$10^{10}L_{\rm \odot}$ \citep{2006gaun.book.....S}, we are well in the limit where $L_{\rm min} \gg L_{*}$ and we can substitute equation~\ref{eq:CminLmin} into equation~\ref{eq:link}, which gives
\begin{equation}\label{eq:constrain}
B_{\rm L,\Delta}/B_{\rm n,\Delta}\approx1.0\;,
%C_{\rm min}= \frac{L_{\rm min}{L_{*}}}\appox 4,
\end{equation}
We checked that our results are not very sensitive to the exact value of $L_{*}$.
In all the following, we fit the data with two models: (1) a model with unconstrained parameters $B_{\rm L,\Delta}$ and $B_{\rm n,\Delta}$, as in \cite{Soumagnac2016} and (2) a more realistic model, reflecting our knowledge of the sample flux limit, where $B_{\rm L,\Delta}=B_{\rm n,\Delta}$ as in equation~\ref{eq:constrain}.
%So converting the limit to absolute mag and then to flux, we obtain 
%L=10**(-(M-4.83)/2.5)/1.4e10 ~ 6 
%(where M=19.9-5*math.log10(3000*1e6)+5, with 3000*1e6 being the distance in pc corresponding to z=0.55 with a standard cosmology), and we took Lstar ~ 1.4 x 10^10 Lsun
%Could you specify where the 2.5 value you suggested come from so that we are on the same page?
%roughly to L = 2.5 L_star. 
%{\color{magenta} Here we need to explain why we set $B_{\rm L,\Delta}=B_{\rm n,\Delta}$. It comes from the fact that the flux limit (you wrote L = 2.5 L_star in your email). The ratio [B_L,Delta / B_n,Delta] should depend only on the flux limit). Wuld be great if you could write a short paragraph on this.}
           
\section{Measurement}\label{sec:data}

\subsection{The BOSS DR12 sample}

In this analysis, we use the public data from the Sloan Digital Sky Survey's (SDSS-III)  Baryon Oscillation Spectroscopic Survey (BOSS), data release 12 (DR1, Alam et al. 2015). The SDSS (York et al. 2000), divided into SDSS I, II (Abazajian et al. 2009), and III (Eisenstein et al. 2011), used a drift-scanning mosaic CCD camera (Gunn et al. 1998) to image over one third of the sky (14,555 square degrees) in five photometric bands [$u,g,r,i,z$] (Fukugita et al. 1996; Doi et al. 2010) to a limiting magnitude of $r \approx$ 22.5 using the dedicated 2.5-m Sloan Telescope located at Apache Point Observatory in New Mexico.

BOSS is primarily a spectroscopic survey, which is designed to obtain spectra and redshifts for $\sim$1.35 million galaxies over an extragalactic footprint covering $\sim$10,000 square degrees. These galaxies are selected from the SDSS DR8 imaging. Together with these galaxies, 160 000 quasars and approximately 100 000 ancillary targets are being observed. The method by which the spectra are obtained (Smee et al. 2013) ensures a homogeneous data set with a high redshift completeness of more than 97$\%$ over the full survey footprint. Redshifts are extracted from the spectra using the methods described in Bolton et al. (2012). A summary of the survey design appears in Eisenstein et al. (2011), and a full description is provided in Dawson et al. (2013).

Two classes of galaxies were selected by BOSS to be targeted for spectroscopy using SDSS DR8 imaging. The ``LOWZ'' algorithm is designed to select red galaxies at $z < 0.45$ from the SDSS DR8 imaging data. While the ``CMASS'' sample is designed to be approximately stellar-mass-limited above z = 0.45. 

In our previous work we considered only the CMASS sample from DR10, in this work we use DR12 data which is $\sim50\%$ larger in angular sky coverage. We leave the analysis using the LOWZ sample for future developments of this work. %We also now consider the LOWZ sample of galaxies, for which the results are presented in the appendix of this paper. 
The details of the catalogue are provided in table \ref{table:data}.

\begin{table}
\centering
\begin{tabular}{lll}
\hline
\multicolumn{1}{|l|}{} &\multicolumn{1}{l|}{CMASS} \\ \hline \hline
redshift range         &  $0.43< z <0.7$           \\
effective redshift     & 0.57                      \\
effective area         & 9376 deg$^2$              \\
effective volume       & 4.70 Gpc$^3$              \\
number of galaxies     & 800,853                  \\ \hline
\end{tabular}
\caption{\label{table:data} Summary of the data samples used. The effective volume is calculated using our fiducial model and the amplitude of the matter power spectrum at the BAO scale $P_0=10,000~h^{-3}$Mpc$^3$. }
\end{table}

\begin{comment}
\begin{table}
\centering
\begin{tabular}{lll}
\hline
\multicolumn{1}{|l|}{} & \multicolumn{1}{l|}{LOWZ} & \multicolumn{1}{l|}{CMASS} \\ \hline \hline
redshift range         & $0.15< z <0.43$           & $0.43< z <0.7$           \\
effective redshift     & 0.32                      & 0.57                      \\
effective area         & 8337 deg$^2$              & 9376 deg$^2$              \\
effective volume       & 2.00 Gpc$^3$              & 4.70 Gpc$^3$              \\
number of galaxies     & 361,775                   & 800,853                  \\ \hline
\end{tabular}
\caption{\label{table:data} Summary of the data samples used. The effective volume is calculated using our fiducial model and the amplitude of the matter power spectrum at the BAO scale $P_0=10,000~h^{-3}$Mpc$^3$. }
\end{table}
\end{comment}

\subsection{Estimator \& Computation}

Several practical problems inhibit our ability to accurately measure the 2PCF of the galaxy distribution, as defined in equation~\ref{eq:xilin}. The discreet sampling by individual galaxies of the smooth density field  leads to shot noise on small scales. Other difficulties arise from the irregular shape of galaxy surveys in angular sky coverage, due to dust extinction, bright stars, tracking of the telescope, etc. We must use statistical estimators which can deal with such problems (see \citet{Percival2007}, for a review of correlation function practicalities and \citealt{Kerscher2000}, for a review of correlation estimators).
In this work, the two-point correlation functions $\xi_n$ and $\xi_L$, are computed using the optimal Landy-Salay estimator \citep{Landy1993} which requires the creation of a catalog of random positions. 

\begin{dmath}\label{eq:landy}
\xi(r)= \frac{DD - 2DR + RR}{RR},
\end{dmath}
where DD, DR and RR represent the number of normalised pairs of points at a particular separation, $r$, between the data (D) and a random catalogue (R).  
%\begin{dmath}
%DD = \frac{1}{\sum w^D (\sum w^D - \bar{w}^D)}\sum_i^{N_D}\sum_{j}^{N_D}\sqcap(|\vec{x}_i^{D}-\vec{x}_j^{D}|)\;,
%\end{dmath}
%\begin{dmath}
%DR = \frac{1}{\sum w^D (\sum w^R - \bar{w}^R)}\sum_i^{N_D}\sum_{j}^{N_R}\sqcap(|\vec{x}_i^{D}-\vec{x}_j^{R}|)\;,
%\end{dmath}
%\begin{dmath}
%DR = \frac{1}{\sum w^R (\sum w^R - \bar{w}^R)}\sum_i^{N_R}\sum_{j}^{N_R}\sqcap(|\vec{x}_i^{R}-\vec{x}_j^{R}|)\;,
%\end{dmath}
%where $N_D$ os the number of data points and $N_R$ is the number of randoms. $\vec{x}^{R}$ and $\vec{x}^{D}$ are the position vectors of the random and data points respectively. Each point has a associated weight, $w$, and $\sqcap$ is the rectangular step function defined as,
%\begin{dmath}
%\sqcap(t) = \begin{cases}
%  0 & \text{if $|t-r| > dr/2$} \\
%  w_i . w_j & \text{if $|t-r| < dr/2$} \\
% \end{cases}
%\end{dmath}

In practice, this computation involves counting of the number of weighted pairs separated by $r$ and normalised by the total number of possible weighted pairs in the galaxy sample, the random sample and the cross counts between galaxy-random points. In our analysis. These counts are computed  using an efficient tree-based, parallel, search algorithm called {\tt KSTAT}\footnote{{\tt KSTAT} is publically available from https://bitbucket.org/csabiu/kstat }. The code is based upon the structure known as ``kd-trees'' which is a way of organizing a set of data in k-dimensional space in such a way that once built, any query requesting a list of points in a neighbourhood can be answered quickly without going through every single point.

\subsection{Measurement of $\xi_n(r)$ and $\xi_L(r)$}\label{sec:lum}

For both the number density correlation function $\xi_n$ and the luminosity-weighted correlation function $\xi_L$, we use the published DR12 data and random catalogs\footnote{\label{published}http://www.sdss3.org/dr10/}, including both the radial FKP weights and the angular systematic weights. 

The FKP weights \citep{1994ApJ...426...23F} are applied to all galaxy and random points according to we assign
to each data point a radial weight of $w_{FKP}=
1/[1 +n(z)P_0]$, where n(z) is the radial number density of galaxies and $P_0$(=10,000) is the amplitude of the power spectrum near the BAO scale. 

Each galaxy is assigned a weight according to,
\begin{equation}
w_i=w_{FKP,i}.w_{sys,i}(w_{rf,i}+w_{fb,i}-1),
\end{equation}
where $w_{sys}$ accounts for the correlations between galaxies and stellar density, while $w_{rf}$ and $w_{fb}$ upweights galaxies according to the missed redshifts of neighbouring targeted galaxies due to {\em redshift failure} and {\em fibre collision}. More details on these weights can be found in \cite{2012MNRAS.424..564R, 2016MNRAS.455.1553R}.
%\subsection{Measurement of $\xi_L(r)$}\label{sec:lum}
%Our measurement of the number density correlation function $\xi_n$, using DR12 positions, is in good agreement with the published\footnote{http://www.sdss3.org/science/boss\_publications.php} DR12 correlation function computed by the BOSS collaboration \citep{Anderson2014}.% as shown in figure~\ref{fig:us_versus_BOSS}. 
The computation of the luminosity-weighted correlation function $\xi_L$ requires several steps which are detailed in the next sections.

\begin{figure*}
\includegraphics[width=8cm]{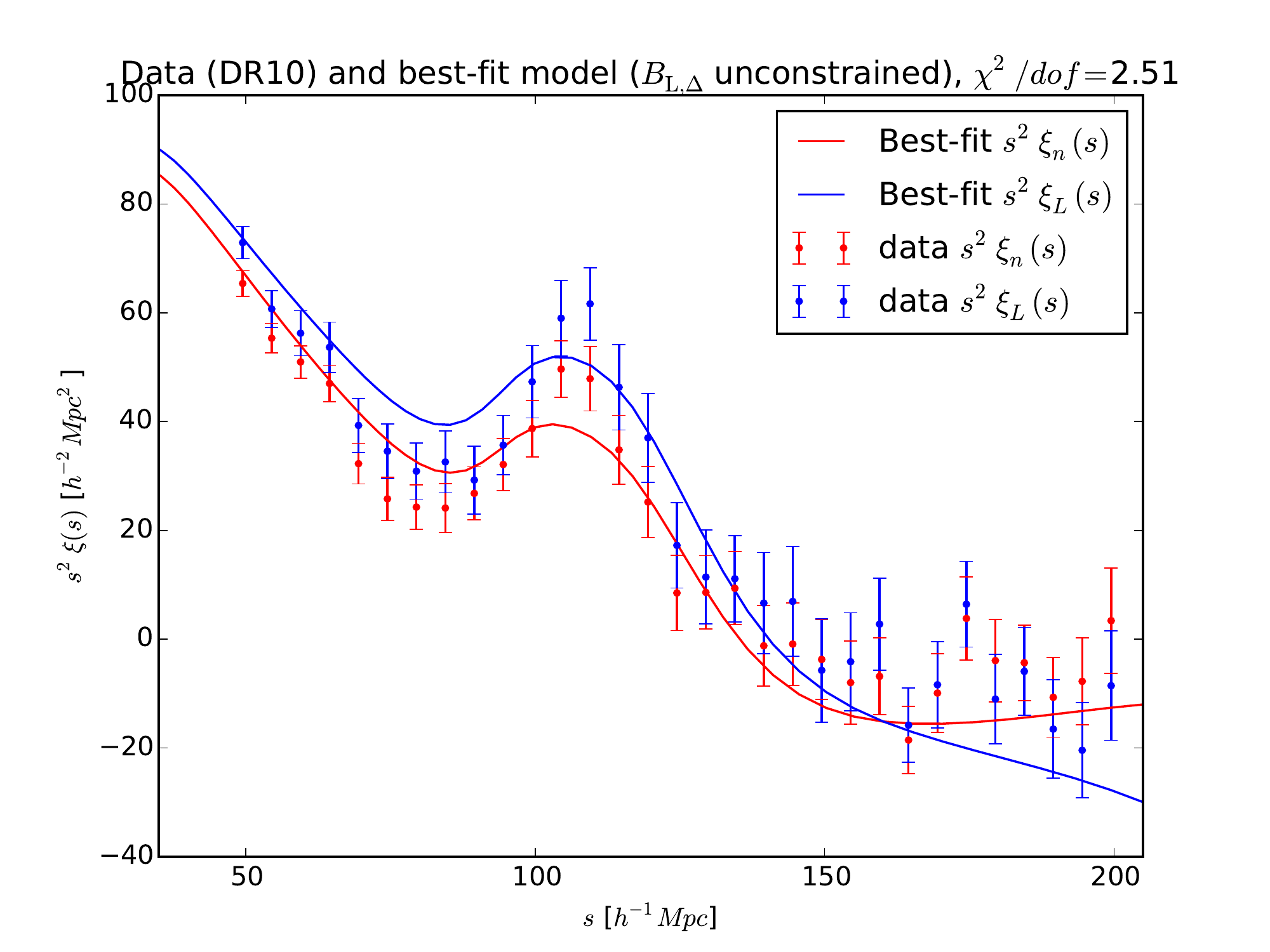}
\includegraphics[width=8cm]{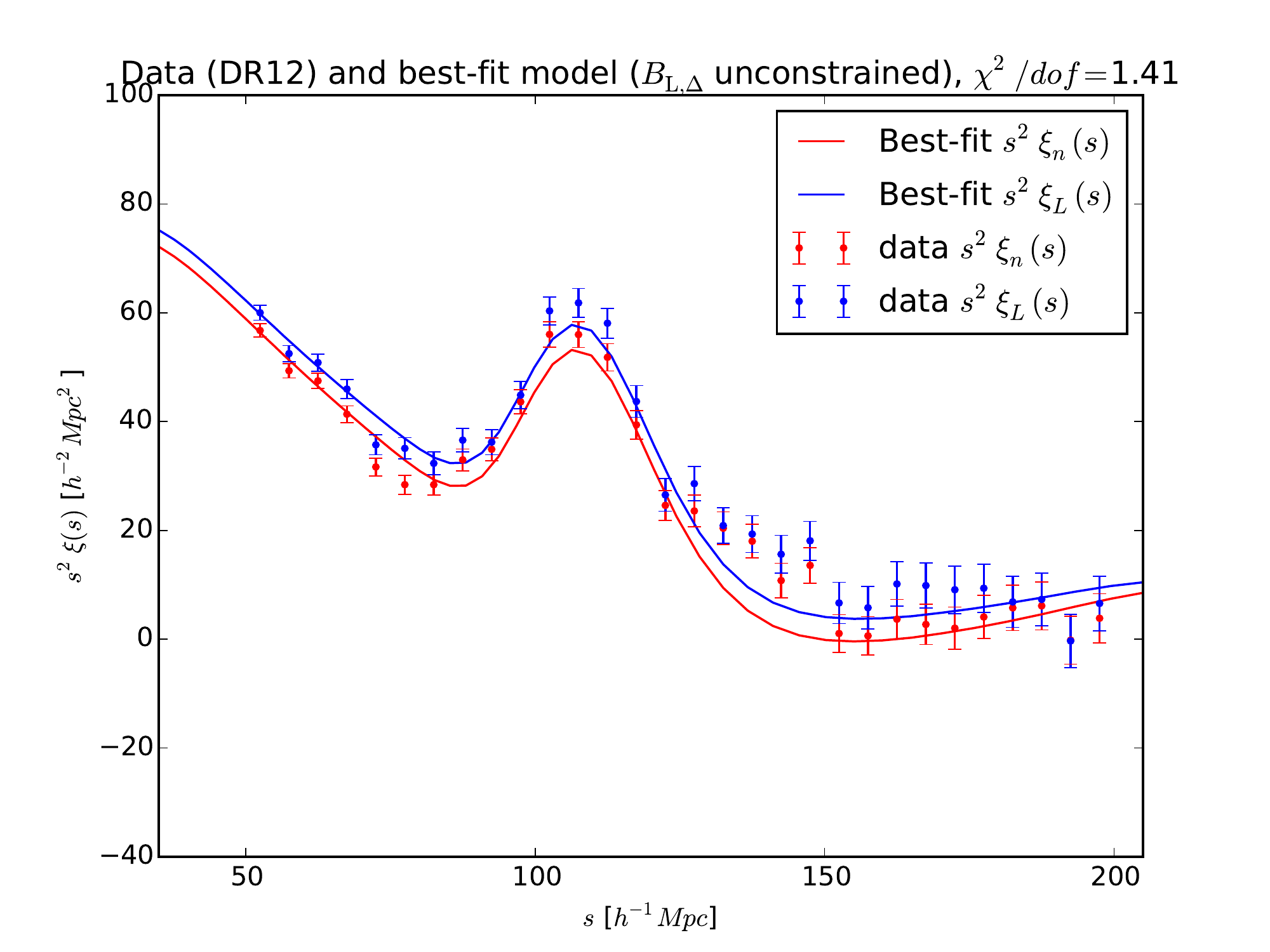}
\includegraphics[width=8cm]{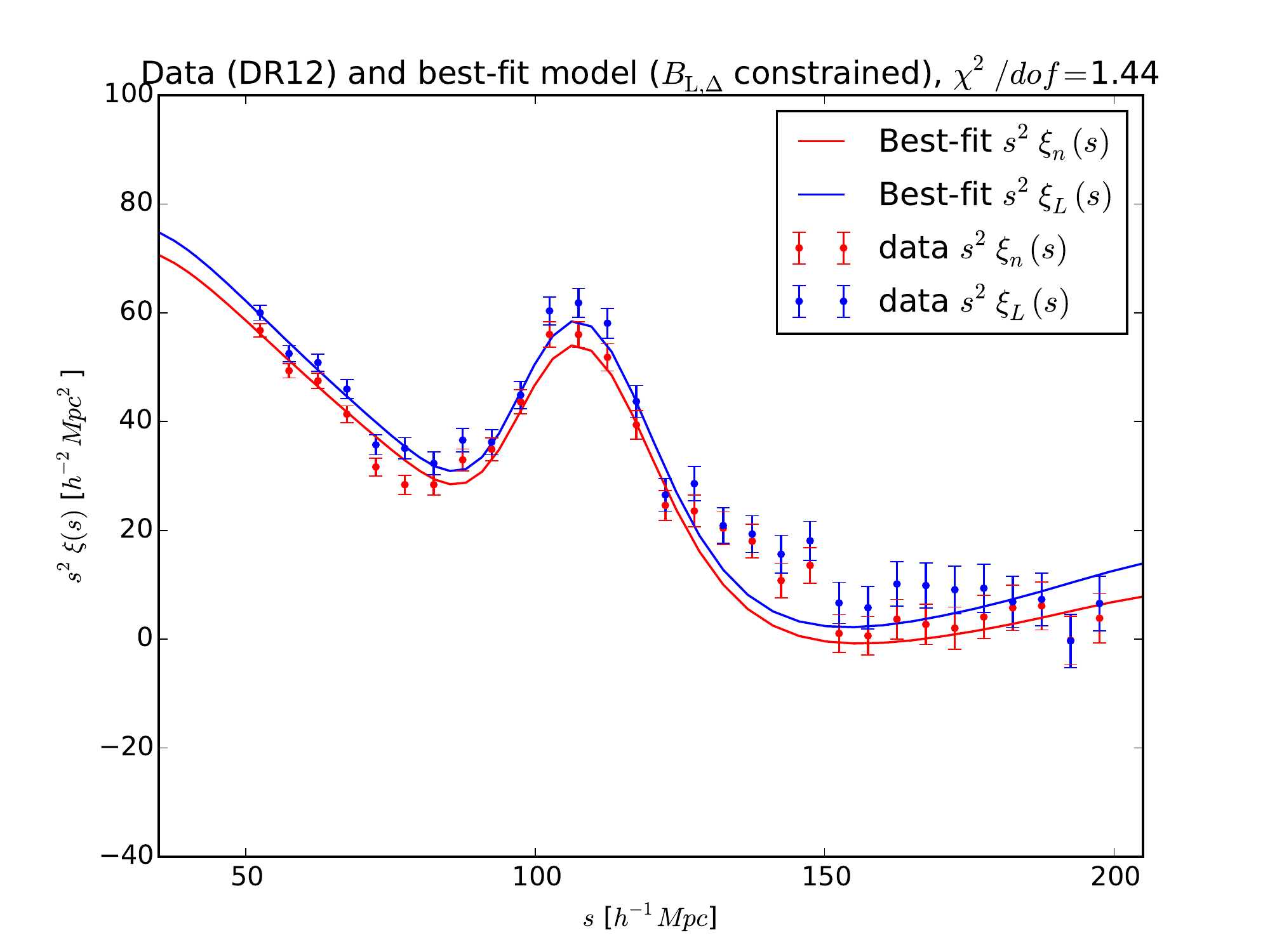}
\caption{Our measurement of the correlation functions $\xi_{\rm n}$ and $\xi_{\rm L}$ (1) with an unconstrained $B_{\rm L,\Delta}$, using the CMASS-DR10 sample (top left) and the CMASS-DR12 sample (top right) and (2) with $B_{\rm L,\Delta}=B_{\rm n,\Delta}$ using the CMASS-DR12 sample (lower panel). A quantitative comparison of all three cases is given in section~\ref{sec:unc} and section~\ref{sec:quantitative_constrained}.}%The figure shows $r^2\cdot\xi_(r)$.}
\label{fig:xi}
\end{figure*}

\subsubsection{Absolute magnitude and absolute luminosity}
We calculate the two-point correlation function of the absolute luminosity density fluctuations, $\xi_L$, using the same estimator and algorithms for $\xi_n$, and weighting each object with its absolute luminosity. The absolute luminosity is calculated using the $i$ and $g$ bands photometric data, from the CMASS DR12 catalogs. We first compute the absolute magnitudes, using a combination of the ``cmodel'' magnitude parameter, referred to as $m_{cm}$, and the extinction parameter, $e$:
\begin{dmath}
M_{\rm abs}=m_{cm}-e-(5log_{10}(D_L)+25) - K_{Corr}\; , 
\end{dmath}
where the luminosity distance $D_L$ (in Mpc) is linked to the comoving distance $D_M$ via $D_L=(1+z)\cdot D_M$. 
The $m_{cm}$ magnitude is a parameter in the DR12 catalogs derived from the composite flux $F_{composite}=f\cdot F_{dev}+(1-f)\cdot F_{exp}$ which is the best fitting linear combination of  the exponential fit and the de Vaucouleurs fit in each band.  

The parameter $e$ encapsulates the extinction correction, i.e. the account for the absorption and scattering of electromagnetic radiation by dust and gas between the observed galaxies and us. It has been computed following \citet{Schlegel1998}. 
%. The galactic extinction corrections in magnitudes at the position of each object

The magnitude is also k-corrected, $K_{Corr}$, to convert the partial flux collected in the given band into the equivalent rest frame band.
 $K_{Corr}$ is obtained from the fitting formulae of \citet{2010MNRAS.405.1409C}.
%In section ? and in Soumagnac16, the methodology relies not only on galaxy positions but also galaxy luminosities. We obtain extinction corrected $i$-band magnitudes from the SDSS skyserver, and proceed to $K$-correct these magnitudes according to \citet{2010MNRAS.405.1409C}. We then convert the magnitudes to luminosities and 

The absolute luminosities are then computed using
\begin{dmath}
L_{\rm abs}=10^{-(M_{\rm abs}-M_{\rm sun})/2.5}\;,
\end{dmath}
where $M_{\rm sun}=4.83$ is the absolute magnitude of the sun. %The distribution for the absolute luminosity in the $i$ and $g$ bands is shown in figure~\ref{fig:abslum}

%EPS
%\begin{figure*}
%\begin{figure}
%\begin{center}
%%\includegraphics[width=6cm]{./figs/Absolute_mag_histo_I}
%%\includegraphics[width=6cm]{./figs/Absolute_mag_histo_G}
%%\includegraphics[width=6cm]{./figs/Absolute_lum_histo_I}
%%\includegraphics[width=6cm]{./figs/Absolute_lum_histo_G}
%\includegraphics[width=6cm]{./figs/Absolute_mag_histo_I.eps}
%\includegraphics[width=6cm]{./figs/Absolute_mag_histo_G.eps}
%\includegraphics[width=6cm]{./figs/Absolute_lum_histo_I.eps}
%\includegraphics[width=6cm]{./figs/Absolute_lum_histo_G.eps}
%\caption{Distribution of the absolute magnitudes (top panels) and absolute luminosity (lower panels) in the North part of CMASS-DR12, using the $i$ band (left panels) and the $g$ band (right panels).} 
%\label{fig:abslum}
%\end{center}
%\end{figure*}
%\end{figure}

\subsubsection{Correlation functions}
Since the published DR12 random catalog does not include the photometric data necessary to compute $\xi_{\rm L}$, we create one by merging the right ascension and declination information from the available random catalog, with the redshift from the data catalog and the absolute luminosity computed from the data catalog, as explained above. 

Our measurements of $\xi_{\rm L}$ and $\xi_{\rm n}$ are shown in figure~\ref{fig:xi}, together with the previous DR10 measurement by \cite{Soumagnac2016} and our best-fit model, as detailed in the next sections. %EPS
\subsection{Covariance matrix}
%\subsubsection{Jackknife sampling}
The fitting procedure that we describe in section~\ref{sec:fit} requires that we estimate the covariance matrix for our measurement. Since the uncertainties of the measurements of $\xi_{\rm n}(r)$ and $\xi_{\rm L}(r)$ at a given point are correlated, we compute the full covariance matrix for the  joint measurement of $\xi_{\rm n}(r)$ and $\xi_{\rm L}(r)$, as in \cite{Soumagnac2016}. 
We use a Jackknife (JK) resampling technique, as in \citet{Scranton2002}. We split the SDSS area into $N_{JK}$ approximately equal area regions (within 10\% error). We then calculate each correlation function removing one area at a time, and generate our full covariance matrix as
\begin{dmath}
C_{ij}=\frac{N-1}{N}\sum^{N_{JK}}_{k=1}\Delta^{k}_{i}\Delta^{k}_{j}, \label{eq:cov}
\end{dmath}
where the sum is over $N_{jk}$ JK samples and,
\begin{dmath}
\Delta^{k}_{i}={[\xi^k(i)-\bar{\xi}(i)]},
\end{dmath}
where $\xi^k(i)$ is the 2PCF of the i-th bin in the k-th JK sample.
We compute the joint covariance matrix for $\xi_{\rm n}$ and for $\xi_{\rm L}$, using 4096 Jackknife samples. This technique differs from the method adopted by the BOSS collaboration, where 600 mock catalogs were produced and used to estimate the covariance matrix for the fit. The mocks are described in \citet{Manera2013} and the full procedure they adopted to compute the covariance matrix is described in \citet{Percival2014}. The reason we adopt a different approach is that we need to calculate the full covariance matrix for the joint measurement of $\xi_{\rm n}$ and $\xi_{\rm L}$. The mock produced by BOSS do not include any photometric information which would allow us to calculate the luminosity-weighted correlation function. In \cite{Soumagnac2016}, the authors used the covariance matrix computed by BOSS as a way to check the consistency of this approach.

%\subsubsection{Covariance matrix for $\xi_{\rm n}$ and $\xi_{\rm L}$}

%, shown in figure~\ref{fig:JK_n} (in figure~\ref{fig:JK_l} respectively). 
%Our covariance and inverse covariance matrix are shown in figure~\ref{fig:covariance}, and compared to the covariance and inverse covariance published by the BOSS collaboration (in the case of $\xi_{\rm n}$). From the comparison with the BOSS-DR12 covariance, it seems like we are slightly over-estimating our covariance, which is not surprising given that the JK is a rather crude method considering the accuracy of the measurement. We show, in section~\ref{sec:fitxin}, that our fit for $\xi_{\rm n}$ is in good agreement with a fit using the BOSS $\xi_{\rm n}$ and covariance matrix, which indicates that this overestimation does not affect the fit. %In table~\ref{paraXin} of section~\ref{sec:fitxin}, we make sure that this overestimation does not affect the fit.

%\begin{figure*}
%\begin{center}
%\includegraphics[width=6cm]{./figs/BOSScov.eps}
%\includegraphics[width=6cm]{./figs/invcovBOSS.eps}
%\includegraphics[width=6cm]{./figs/cov_xinonly_dr10.eps}
%\includegraphics[width=6cm]{./figs/invcov_xinonly_dr10.eps}
%\caption{Covariance matrix (left) and inverse covariance matrix (right) for the $\xi_{\rm n}$ measurement, measured by the BOSS collaboration (upper panel; published DR11 covariance matrix), and by us (lower pannel)}
%\label{fig:covariance}
%\end{center}
%\end{figure*}

\begin{figure}
\begin{center}
\includegraphics[width=8cm]{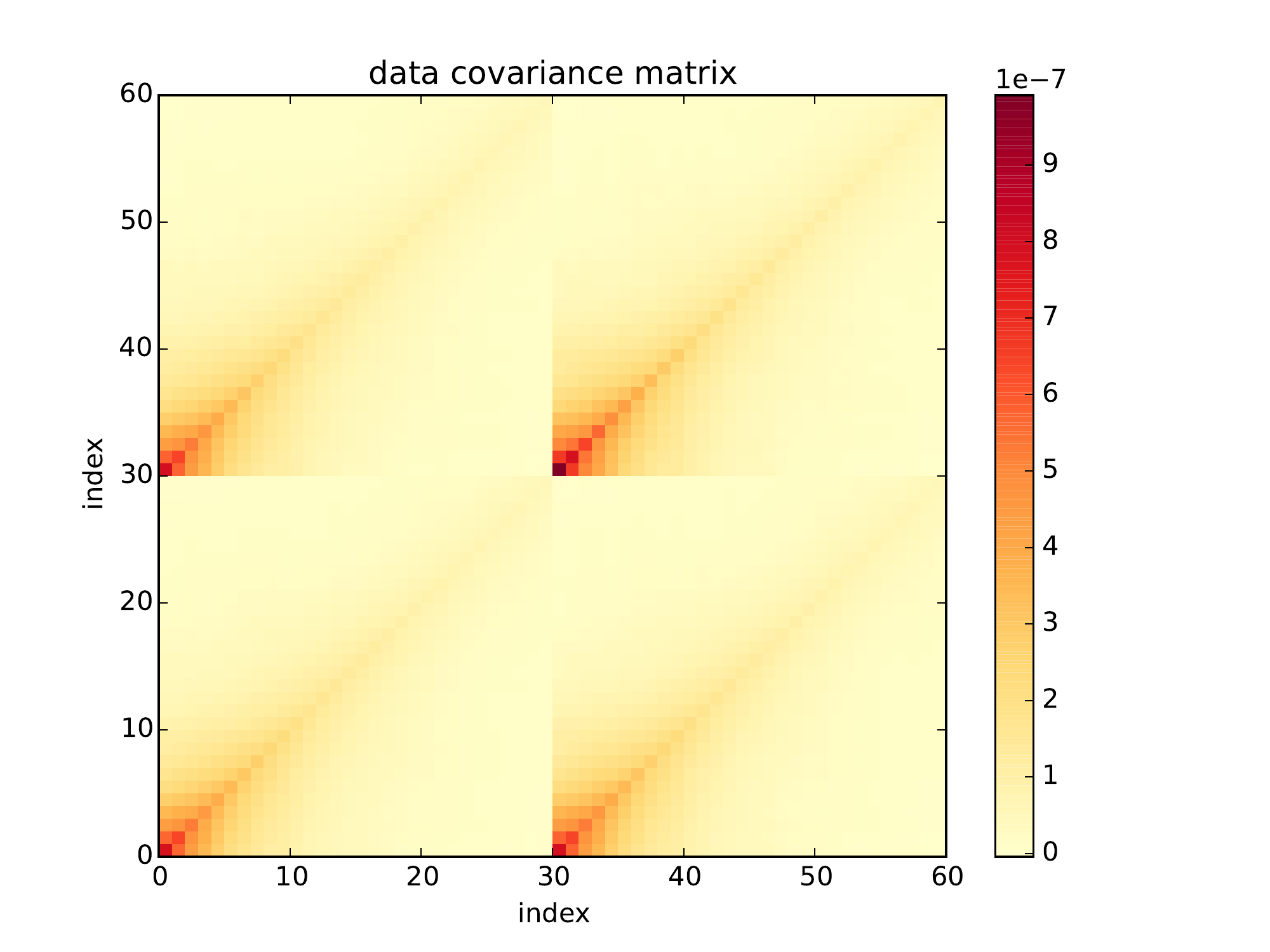}
\caption{Our measurement of the joint covariance matrix using the CMASS-DR12 sample. It is not diagonal: the uncertainties on $\xi_{\rm n}$ and $\xi_{\rm L}$ are correlated, which underlines the importance of performing a joint fit of $\xi_{\rm n}$ and $\xi_{\rm L}$.}%which leads us to perform a joint fit of $\xi_{\rm n}$ and $\xi_{\rm L}$ rather than separate fits.}
\label{fig:covariance_nl}
\end{center}
\end{figure}

%\subsubsection{Joint covariance matrix for $\xi_{\rm n}$ and $\xi_{\rm L}$}
The full covariance matrix is shown in figure~\ref{fig:covariance_nl}. It is far from being diagonal, or even block-diagonal, which shows the importance of fitting $\xi_{\rm n}$ and $\xi_{\rm L}$ jointly. 

%TO DO : ADD THE 21 BINS COVARIANCES

\section{Results}
We explored the two different cases presented in section~\ref{sec:previous_results} and section ~\ref{sec:constrained}, namely a model with an unconstrained parameter $B_{\rm{L,\Delta}}$ and a model with a more realistic constraint $B_{\rm{L,\Delta}}=B_{\rm{n,\Delta}}$. We applied to both cases the methodology developed in \cite{Soumagnac2016} - which we present here in further details - to determine whether we detect a scale-dependent bias of the luminosity correlation function in the DR12 data, i.e. a non zero value of $B_{\rm L,\Delta}$.
%We explored the model presented in section~\ref{sec:previous_results}, where $B_{\rm{L,\Delta}}=1.1B_{\rm{n,\Delta}}$ and 
%We apply the methodology developed in \cite{Soumagnac2016} and presented here in further details, to determine wether we detect a scale-dependent bias of the luminosity correlation function in the DR12 data, i.e. a non zero value of $B_{\rm{L,\Delta}}=1.1B_{\rm{n,\Delta}}$.
%In section FIT
%To determine whether we detect a scale-dependent bias of the
%luminosity correlation function requires answering the following
%question: do the data support the inclusion of a non-zero extra
%parameter $B_{\rm L,\Delta}$? Rather than a question of parameter
%estimation, this is a question of model comparison between two models
%$\mathcal{M}$, with or without $B_{\rm
  %L,\Delta}$. We present the formalism and results of ur model comparision in section ADD

\subsection{Model Fitting}\label{sec:fit}
\subsubsection{Formalism and computation}
%This section focuses on the strategy we adopt to fit the model presented in section~\ref{sec:mod} to the measurement presented in section~\ref{sec:data}.
%We present the model fitting basic formalism in section~\ref{sec:modelfit}, and the algorithm we used to perform Monte Carlo Markov Chains (MCMC) in section~\ref{sec:emcee}. 
%We show our results, first when fitting only $\xi_{\rm n}$, in sections~\ref{sec:fitxin}, and then when fitting jointly $\xi_{\rm n}$ and $\xi_{\rm L}$, in section~\ref{sec:fitxinxil}. 

%\subsection{Formalism}\label{sec:modelfit}

We adopt the terminology of \citet{Hogg2010}, defining a {\it generative model} (a parametrized quantitative description of a statistical procedure that could reasonably have generated the data) and an {\it objective scalar} to be optimized. 
We assume that the only reason that our data point deviate from the model described by equations~\ref{eq:xin_new} and~\ref{eq:xil} is an offset in the $\xi$ direction, drawn from a gaussian distribution of zero mean and known variances $\sigma_{\xi}$. We wish to get the set of parameters $\theta=\{B_{\rm n,t},B_{\rm
  L,t},B_{\rm n,\Delta},B_{\rm CIP},B_{\rm
  sys,n},B_{\rm sys,L},A_{\rm MC},k_*\}$ which maximizes the probability of our model $\mathcal{M}$ given the data $\mathcal{D}$, i.e. the posterior probability distribution $Pr(\theta|\{\mathcal{D},\mathcal{M}\})$. Bayes' theorem relates it to the likelihood $\mathcal{L}\equiv Pr(\mathcal{D} | \theta,\mathcal{M})$, via the prior $\pi\equiv Pr(\mathcal{D}|\mathcal{M},\theta)$:
\begin{equation}
Pr(\theta|\{\mathcal{D},\mathcal{M}\})=\frac{Pr(\mathcal{D}|\{\theta,\mathcal{M}\})\cdot Pr(\theta|\mathcal{M})}{Pr(\mathcal{D}|\mathcal{M})}
=\frac{\mathcal{L}\cdot \pi}{E}\;,
\end{equation}
where the evidence $E=Pr(\mathcal{D}|\mathcal{M})$ is the probability of getting the data $\mathcal{D}$, given the model $\mathcal{M}$ and can be seen as the likelihood averaged over all the possible parameters within a model.

Within the framework of a model-fitting approach, the evidence is seen as a marginalization constant and is ignored, since it does not affect the result of the optimization of the objective scalar. This is no longer true when adopting a model selection approach to our problem, as will be discussed in section~\ref{sec:sel}.
The likelihood of our generative model is :
\begin{dmath}
\mathcal{L} \propto \exp\left[-\frac{1}{2}R^T\cdot C^{-1}\cdot R\right]
\end{dmath}
where ${\bf R}={\bf Y}-{\bf AX}$, and $C^{-1}$ is the inverse covariance matrix of the data ${\bf Y}$. 
%\subsubsection{Priors}
We apply the following uniform (not ``informative'') priors for the nine parameters of our model (the same as in \citealt{Soumagnac2016}):
\begin{itemize}
\item $B_{\rm n,t}\in [0,5]$
\item $B_{\rm L,t}\in [0,5]$
\item $B_{\rm n,\Delta}\in[-10,10]$
\item $B_{\rm sys,L} \in [-0.01,0.01]$
\item $B_{\rm sys,n} \in [-0.0015,0.0015]$
\item $k_{*} \in [0,10]$
\item $A_{\rm MC} \in [0,6]$
\item $B_{\rm CIP} \in [-0.3,0.3]$
\item $B_{\rm L,\Delta}\in[-10,10]$ (only in the case described in section~\ref{sec:unc})
\end{itemize}
%For the parameter $B_{\rm L,\Delta}$, in the case where it is unconstrained we apply the priopr $B_{\rm L,\Delta} \in [-10,10]$.

We believe this is a conservative choice of priors. The priors on
$B_{\rm L,\Delta}$ and $B_{\rm CIP}$ are intentionally taken to be
broad. The priors on $B_{\rm sys,L}$ and
$B_{\rm sys,n}$ are based on a study \citep{Ross2012} of the potential
systematic effects in the BOSS data; this limit effectively allows a
systematic contribution that is up to 3 times as large as the
systematic contribution to $\xi_{\rm n}$ found in BOSS.  The priors on
$B_{\rm n,t}$, $k_*$ and $A_{\rm MC}$ are taken to be consistent with
previous works on the BOSS data \citep{Crocce2008,Anderson2012,Anderson2014}.

%{\color{red} change this} The parameters $C$ and $D$
%depend on the flux threshold of the sample; given that the measured
%luminosity function of the CMASS DR10 sample extends well into the
%faint end (within a Schechter luminosity function fit), we expect $C,D
%\approx ?$

%We believe this is a conservative choice of priors. The prior on $b_4$ is willingly taken to be broad, although BL11 forecasted it to be around $b_4\approx2.6$ (in a case where $C_{\rm min}=D_{\rm min}=0$, $b_{\rm n}=2$ and $b_{\rm L;t}=1$). The prior on $b_{CIP}$ is taken to be broader than the upper limit of $10^{-2}$ set in \citet{Grin2014} for $A_{CIP}=b_{CIP}/b_4^2$. The prior on $b_{sys}$ is based on the study by \cite{Ross2012} of the potential systematic effects in the BOSS data: it is taken to be $\sim3$ times the size of the systematic contribution to the BOSS $\xi_{\rm n}$ (without systematic weight). The priors on $b_1$, $k_*$ and $A_{\rm MC}$ are taken to be consistent with previous works on the BOSS data \citep{Crocce2008, Anderson2012,Anderson2014}. The prior on $b_2$ is also chosen to be broad, although preliminary estimations of this parameter that we made, indicate it should be close to zero in the case of the CMASS DR12 sample.

%\subsection{Computation}\label{sec:emcee}

%\subsubsection{Computation}
In the case of a not informative prior, the optimisation of the likelihood function corresponds to the maximum of the posterior distribution, i.e. the maximum a posteriori value. The problem then becomes to estimate the uncertainties on the maximum a posteriori value of each parameter, i.e. obtain the distribution of parameters that is consistent with our data, and to be able to marginalise over it to get the distribution of each parameter. This is made possible by Monte Carlo Markov Chain (MCMC) sampling.
%with the {\it scipy.optimize} python modul.
We used the multimodal nested sampling algorithm, {\it MultiNest}
\citep{Feroz2008} to sample from the posterior probability distribution, and quote the uncertainties based on the $16^{th}$, $50^{th}$, and $84^{th}$ percentiles of the samples in the marginalised distributions, corresponding to $1\sigma$ in the case of a gaussian.

\begin{figure*}
\begin{center}
\includegraphics[width=8cm]{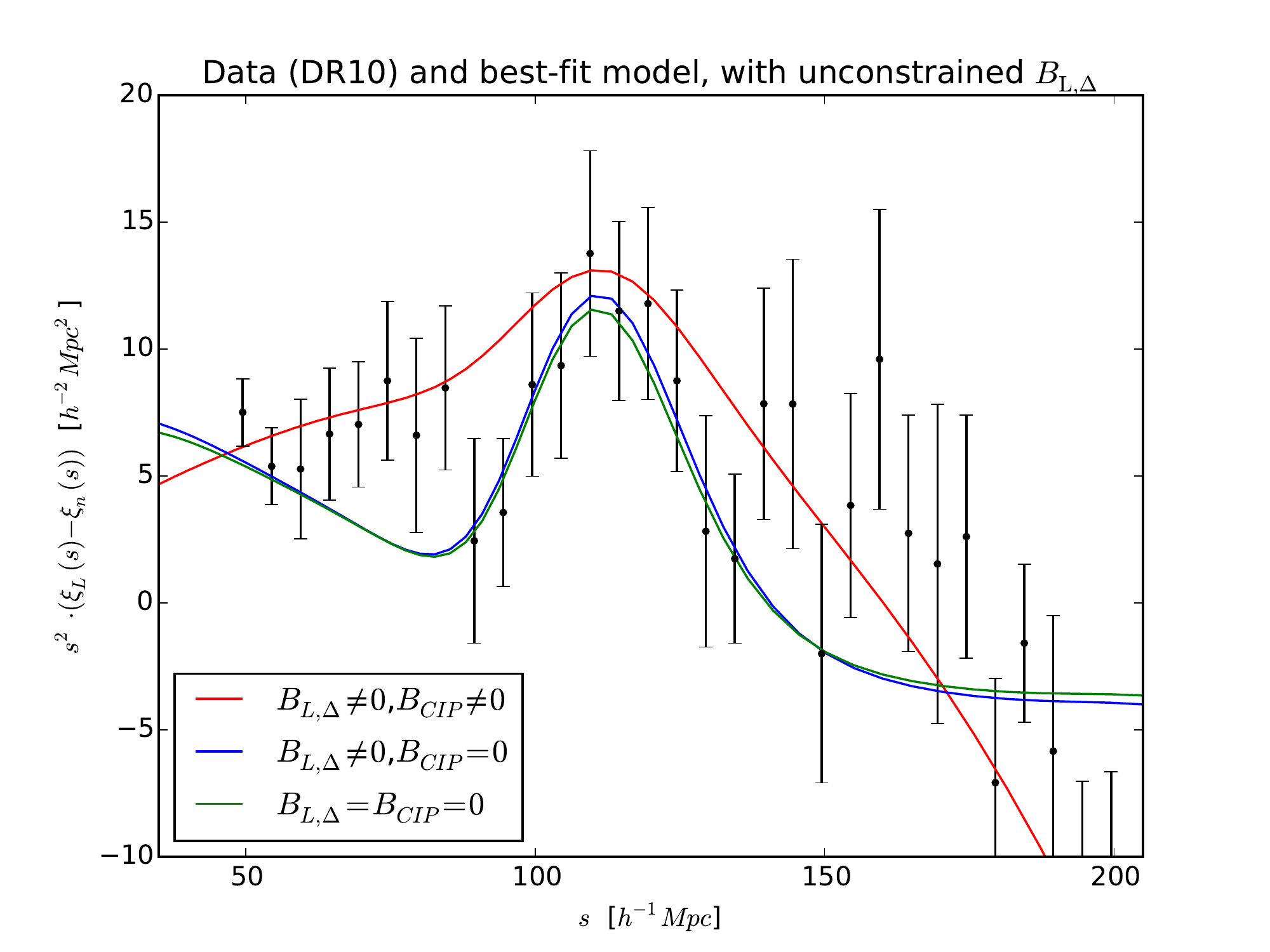}
\includegraphics[width=8cm]{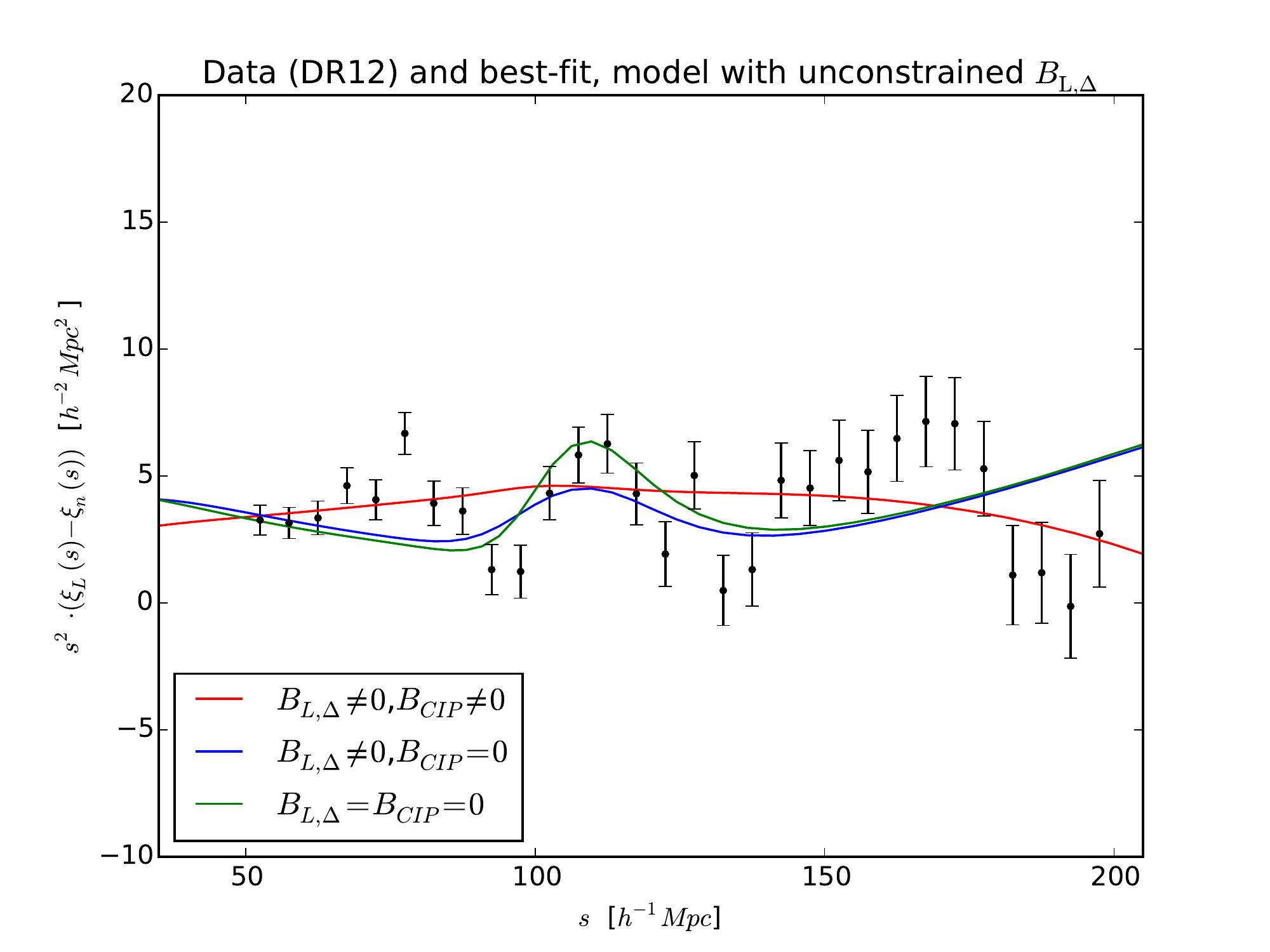}
\includegraphics[width=8cm]{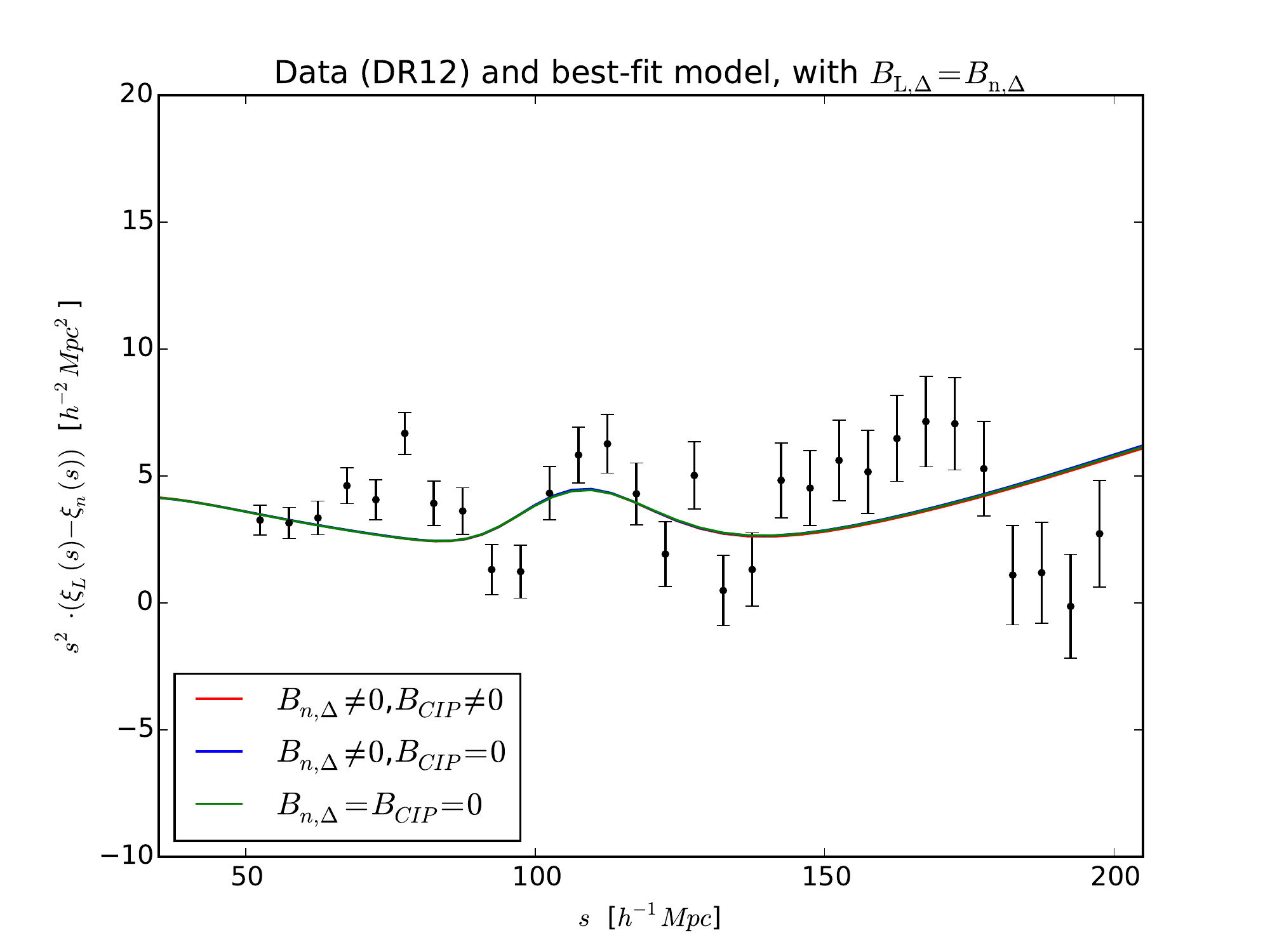}
\caption{Our measurement of the difference $\xi_{\rm L}-\xi_{\rm n}$
  (times $s^2$). The red line corresponds to our full model, the
  blue line corresponds to a model with $B_{\rm CIP}=0$, and the green
  line corresponds to a model with $B_{\rm CIP}=B_{\rm L,\Delta}=0$. The top panels show the data and best fit in the unconstrained case, using the CMASS-DR10 sample as in Soumagnac et al. 2016 (left) and the CMASS-DR12 sample (right). The lower panel shows the more realistic case where $B_{\rm L,\Delta}=B_{\rm n,\Delta}$, using the CMASS-DR12 sample. A quantitative comparison of all three cases is given in section~\ref{sec:unc} and section~\ref{sec:quantitative_constrained}.}%which leads us to perform a joint fit of $\xi_{\rm n}$ and $\xi_{\rm L}$ rather than separate fits.}
\label{fig:diff}
\end{center}
\end{figure*}

We consider two cases, corresponding to the presence or absence of
CIPs. %the resulting parameter values are shown in Table~\ref{tab:table3}.
In Figures~\ref{fig:xi} and~\ref{fig:diff}, we show the data and best
fits for the correlation functions $r^2\xi_{\rm n}$ and $r^2\xi_{\rm
  L}$, and for a key quantity, their difference $r^2(\xi_{\rm
  L}-\xi_{\rm n})$. 

%\subsection{Results}
%\begin{comment}
\subsubsection{Unconstrained $B_{\rm{L,\Delta}}$}\label{sec:unc}
\begin{figure*}
\begin{center}
\includegraphics[width=15cm]{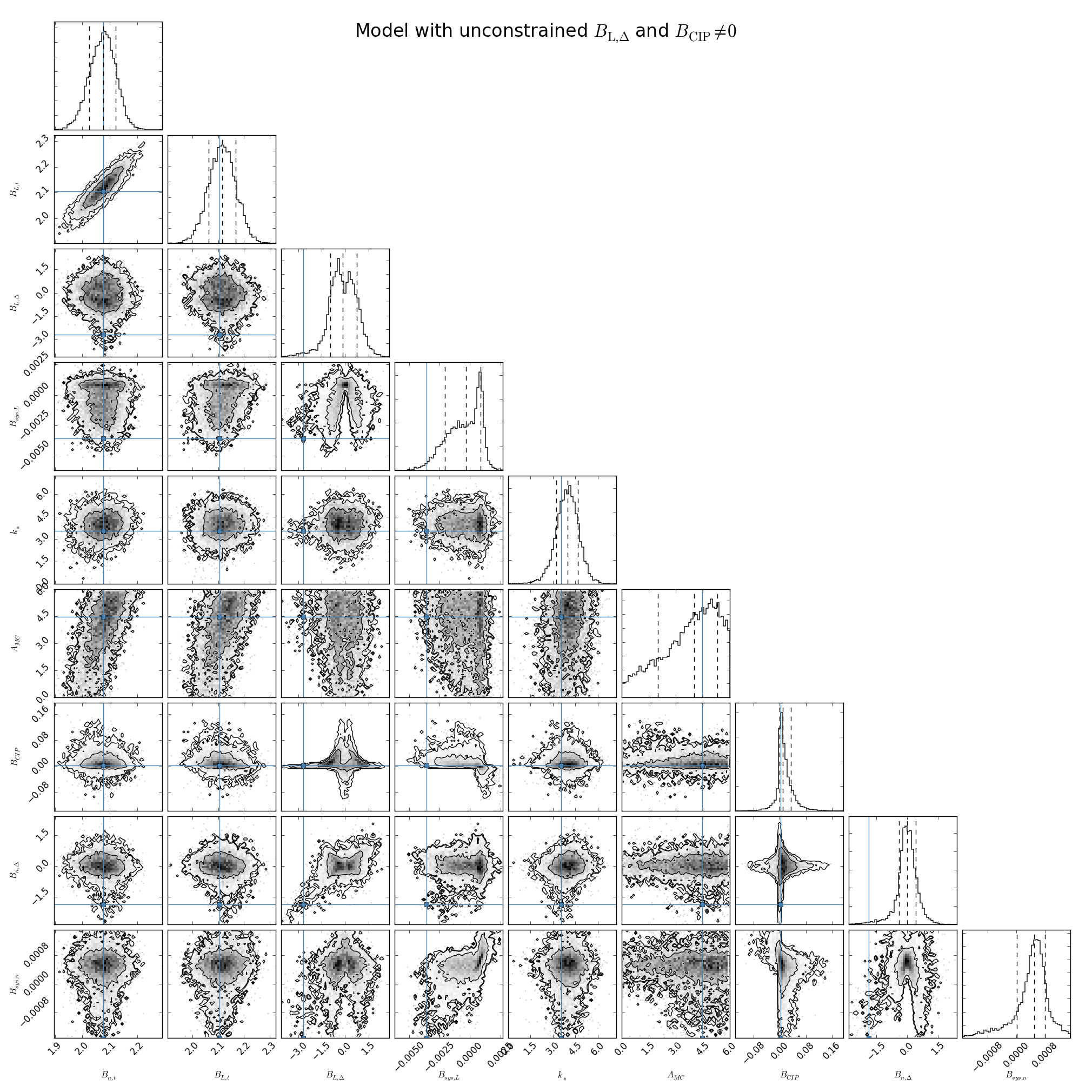}
%this figure is a copy of the one in project/BAO/pdf_northsouth_goodcontours
\caption{{\bf Joint fit of $\xi_{\rm n}$ and $\xi_{\rm L}$, unconstrained $B_{\rm L,\Delta}$ and $B_{\rm CIP}\neq0$:} all the one and two dimensional projections of the posterior probability distributions of the parameters,\{$B_{\rm n,t}$,$B_{\rm
  L,t}$,$B_{\rm n,\Delta}$,$B_{\rm L,\Delta}$,$B_{\rm CIP}$,$B_{\rm sys,n}$,$B_{\rm sys,L}$,$A_{\rm MC}$,$k_*$\}, in the case where $B_{\rm L,\Delta}$ is unconstrained, using the CMASS-DR12 sample. This quickly demonstrates all of the covariances between parameters.The contours correspond to the $1\sigma$, $2\sigma$ and $3\sigma$ percentiles. The blue line corresponds to the maximum likelihood value of each parameter, which is also the maximum a posteriori value (m.a.p.). The dashed lines show the $1\sigma$ percentile of the marginalized distributions.}
\label{fig:triangle_unconstrained}
\end{center}
\end{figure*}

\begin{figure*}
\begin{center}
\includegraphics[width=15cm]{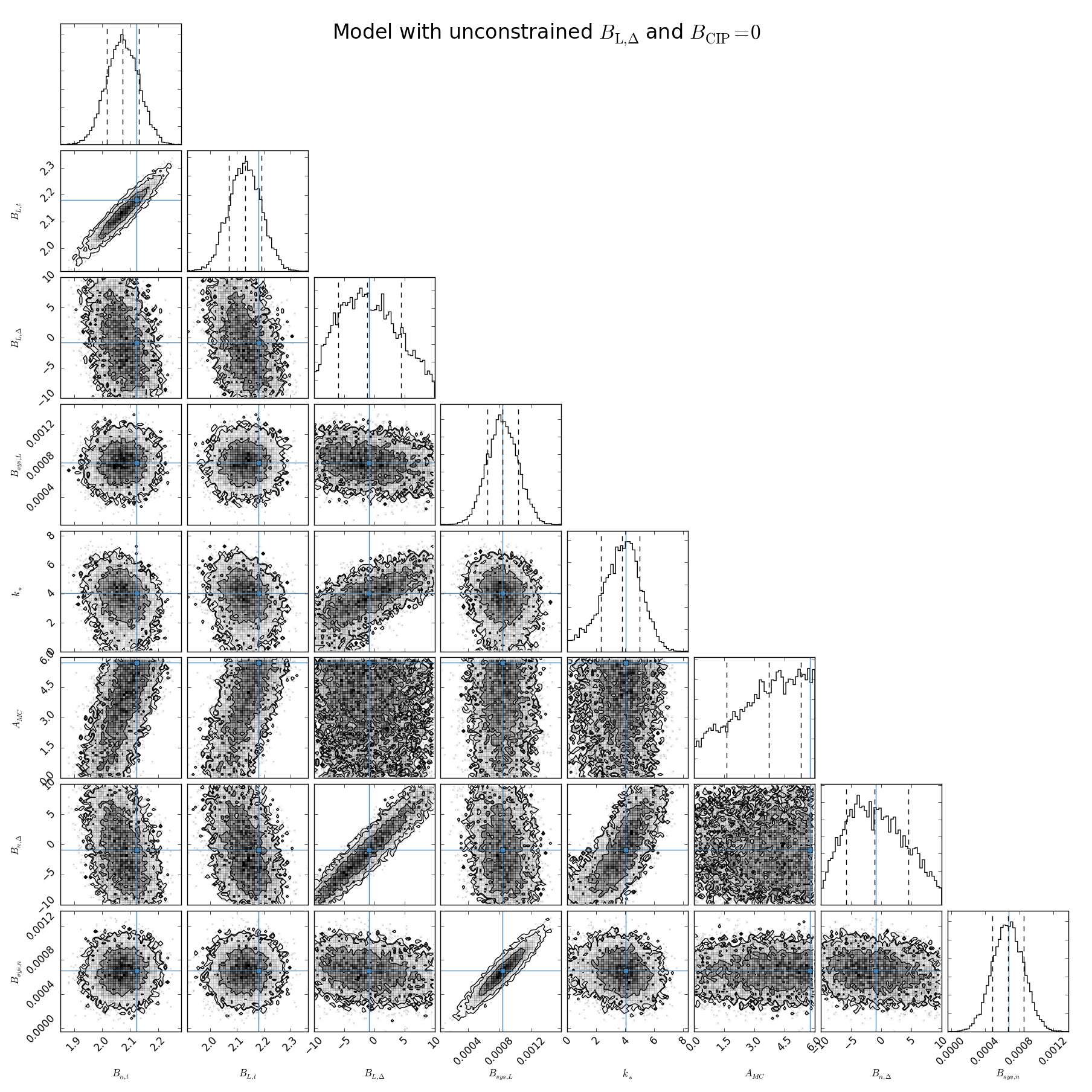}
%this figure is a copy of the one in project/BAO/pdf_northsouth_goodcontours
\caption{{\bf Joint fit of $\xi_{\rm n}$ and $\xi_{\rm L}$, unconstrained $B_{\rm L,\Delta}$ and $B_{\rm CIP}=0$:} all the one and two dimensional projections of the posterior probability distributions of the parameters,\{$B_{\rm n,t}$,$B_{\rm
  L,t}$,$B_{\rm n,\Delta}$,$B_{\rm L,\Delta}$,$B_{\rm CIP}$,$B_{\rm sys,n}$,$B_{\rm sys,L}$,$A_{\rm MC}$,$k_*$\}, in the case where $B_{\rm L,\Delta}$ is unconstrained, using the CMASS-DR12 sample. This quickly demonstrates all of the covariances between parameters.The contours correspond to the $1\sigma$, $2\sigma$ and $3\sigma$ percentiles. The blue line corresponds to the maximum likelihood value of each parameter, which is also the maximum a posteriori value (m.a.p.). The dashed lines show the $1\sigma$ percentile of the marginalized distributions.}
\label{fig:triangle_unconstrained_BCIPnull}
\end{center}
\end{figure*}

In figure~\ref{fig:triangle_unconstrained} and figure~\ref{fig:triangle_unconstrained_BCIPnull}, we show the two dimensional projections of the posterior probability distributions obtained by fitting the model from \cite{Soumagnac2016} (i.e. with unconstrained $B_{\rm{L,\Delta}}$) to the DR12 CMASS data. More specifically, Figure~\ref{fig:triangle_unconstrained} corresponds to the case $B_{\rm CIP}\neq 0$ and figure~\ref{fig:triangle_unconstrained_BCIPnull} to the case $B_{\rm CIP}=0$. The value corresponding to the $16^{th}$, $50^{th}$, and $84^{th}$ percentiles of the samples in the marginalised distributions, i.e. the median value and the $1\sigma$ values (in the case of a gaussian) values are shown in table~\ref{table:paraXinXil_unconstrained}. The best fits to the data are shown and can be compared to the DR10 result in the top left panel of figure~\ref{fig:diff}.

When CIPs are included in the model, i.e. when $B_{\rm CIP}\neq0$, the $1\sigma$ range of $-0.91<B_{\rm L,\Delta}<0.74$ is consistent with zero, and in tension with the prediction of BL11 of $B_{\rm L,\Delta} \approx 2.6$ predicted along with the expectations of $B_{\rm
      n,\Delta} \approx 0$ (this assumption is likely not verified here, as discussed in section ~\ref{sec:constrained}, but this does not affect our result since we did not assume it), and $B_{\rm n,t}$ and $B_{\rm L,t}$ approximately equal. This is an important difference from the result of \cite{Soumagnac2016}, where the authors obtained evidence at $3.2\sigma$ of $B_{\rm L,\Delta}>0.4$ (and evidence that $|B_{\rm L,\Delta}|>0.4$ at $3.7\sigma$) when allowing CIPs, which was an indication of the effect we search for.
%(our maximum likelihood value is ${\color{red}add}$) In BL11, the $2.6$
      In addition, our best-fit value of $B_{\rm CIP}$ is $2.7\times 10^{-3}$, with a $1\sigma$ upper limit
  of % 21 bins $B_{\rm CIP}=$B_{\rm CIP}=6.76\times 10^{-3} and 2 sigma limit is $3.18\times 10^{-2}$,
  $B_{\rm CIP}=3.6\times 10^{-2}$, similar to the $3.7\times 10^{-2}$ upper limit provided by the DR10 data in \cite{Soumagnac2016}\footnote{see supplemental material} and within an order of magnitude of the best existing limits noted previously. 
   
In the absence of CIPs, i.e. when $B_{\rm CIP}$ is set to zero,  $B_{\rm L,\Delta}$ is less constrained. In this case, the $1\sigma$ range $-5.8<B_{\rm L,\Delta}<4.6$ is consistent with both zero and the BL11 prediction and similar to the $[-2.8, 7.6]$ range computed with the DR10 data.

%\end{comment}

\subsubsection{More realistic model with $B_{\rm{L,\Delta}}=B_{\rm{n,\Delta}}$}\label{sec:quantitative_constrained}

\begin{figure*}
\begin{center}
\includegraphics[width=15cm]{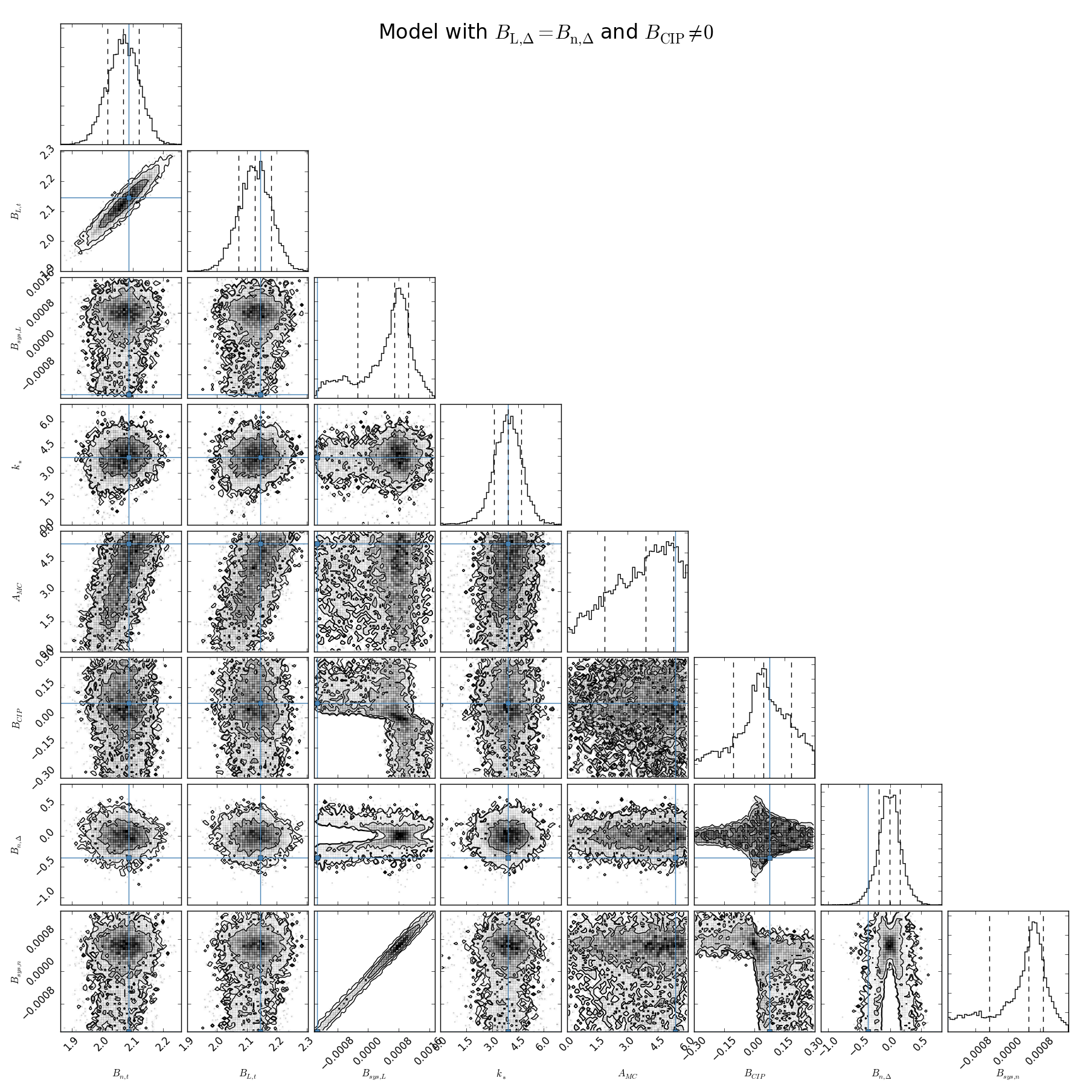}
%this figure is a copy of the one in project/BAO/pdf_northsouth_goodcontours
\caption{{\bf Joint fit of $\xi_{\rm n}$ and $\xi_{\rm L}$, with the constrain $B_{\rm L,\Delta}=B_{\rm n,Delta}$ and $B_{\rm CIP}\neq0$:} all the one and two dimensional projections of the posterior probability distributions of the parameters,\{$B_{\rm n,t}$,$B_{\rm
  L,t}$,$B_{\rm n,\Delta}$,$B_{\rm CIP}$,$B_{\rm sys,n}$,$B_{\rm sys,L}$,$A_{\rm MC}$,$k_*$\}, in the case where $B_{\rm L,\Delta}=B_{\rm n,\Delta}$ is unconstrained, using the CMASS-DR12 sample. This quickly demonstrates all of the covariances between parameters.The contours correspond to the $1\sigma$, $2\sigma$ and $3\sigma$ percentiles. The blue line corresponds to the maximum likelihood value of each parameter, which is also the maximum a posteriori value (m.a.p.). The dashed lines show the $1\sigma$ percentile of the marginalized distributions.}
\label{fig:triangle_constrained}
\end{center}
\end{figure*}

\begin{figure*}
\begin{center}
\includegraphics[width=15cm]{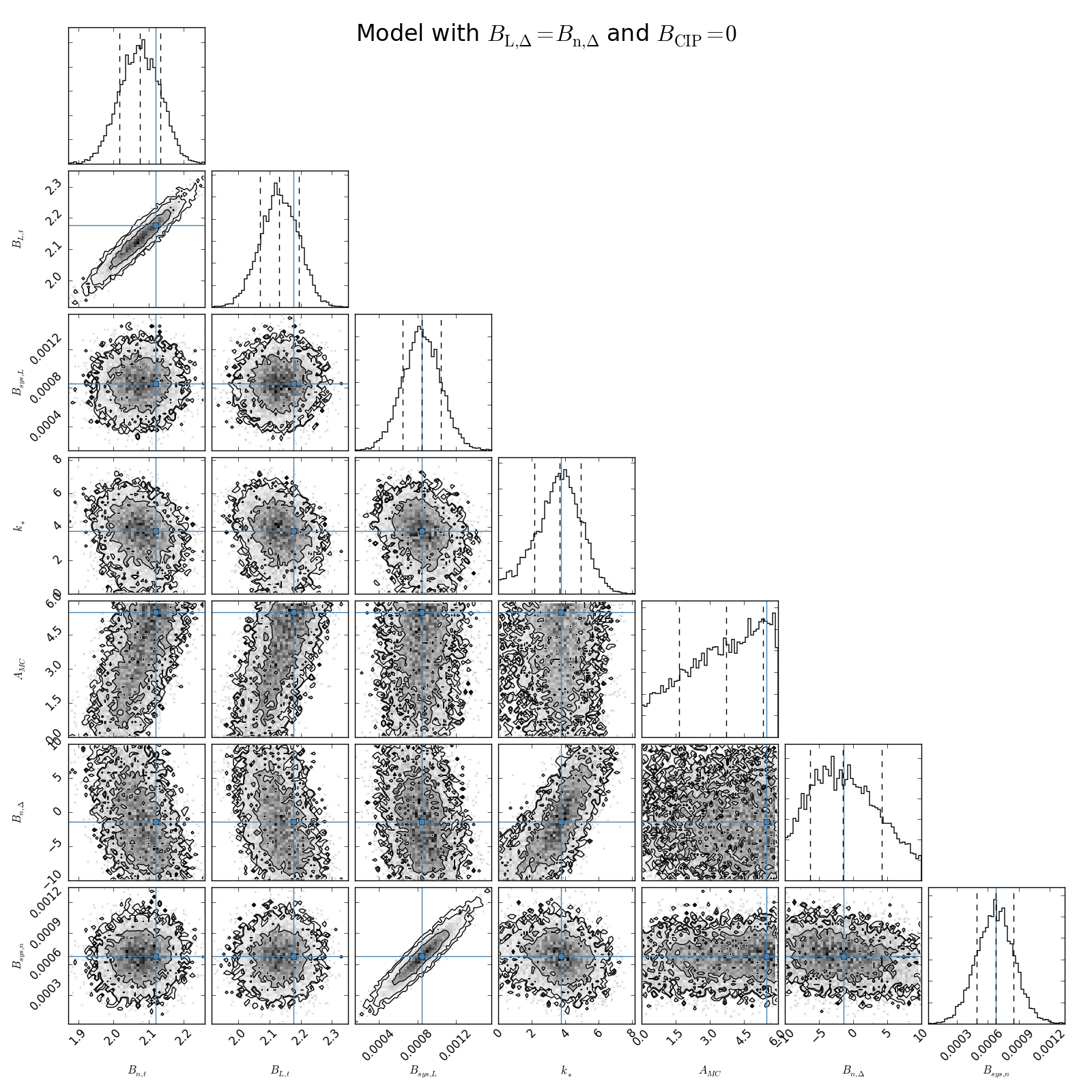}
%this figure is a copy of the one in project/BAO/pdf_northsouth_goodcontours
\caption{{\bf Joint fit of $\xi_{\rm n}$ and $\xi_{\rm L}$, with the constrain $B_{\rm L,\Delta}=B_{\rm n,Delta}$ and $B_{\rm CIP}=0$:} all the one and two dimensional projections of the posterior probability distributions of the parameters,\{$B_{\rm n,t}$,$B_{\rm
  L,t}$,$B_{\rm n,\Delta}$,$B_{\rm CIP}$,$B_{\rm sys,n}$,$B_{\rm sys,L}$,$A_{\rm MC}$,$k_*$\}, in the case where $B_{\rm L,\Delta}=B_{\rm n,\Delta}$ is unconstrained, using the CMASS-DR12 sample. This quickly demonstrates all of the covariances between parameters.The contours correspond to the $1\sigma$, $2\sigma$ and $3\sigma$ percentiles. The blue line corresponds to the maximum likelihood value of each parameter, which is also the maximum a posteriori value (m.a.p.). The dashed lines show the $1\sigma$ percentile of the marginalized distributions.}
\label{fig:triangle_constrained_BCIPnull}
\end{center}
\end{figure*}

In figure~\ref{fig:triangle_constrained} and figure~\ref{fig:triangle_constrained_BCIPnull}, we show the two dimensional projections of the posterior probability distributions obtained by fitting a model where $B_{\rm L,\Delta}=B_{\rm n,\Delta}$ to the DR12 CMASS data. Figure~\ref{fig:triangle_constrained} corresponds to the case $B_{\rm CIP}\neq 0$ and figure~\ref{fig:triangle_constrained_BCIPnull} to the case $B_{\rm CIP}=0$. A full
  tabulation of our best-fit parameters is shown in table~\ref{table:paraXinXil_constrained}. The best fits to the data are shown in figure~\ref{fig:xi} and figure~\ref{fig:diff} and can be compared to the DR10 result.
  
When CIPs are included in the model, i.e. when $B_{\rm CIP}\neq0$, the $1\sigma$ range $-0.183<B_{\rm n,\Delta}<0.168$ is consistent with zero, and is in tension with the prediction in BL11 of $B_{\rm L,\Delta} \approx 2.6$, predicted along with the expectations of $B_{\rm n,\Delta} \approx 0$, (this latter assumption is likely not verified here, as discussed in section~\ref{sec:constrained}) and $B_{\rm n,t}$ and $B_{\rm L,t}$ approximately equal (which is indeed included in our model through the constraint $B_{\rm L,\Delta}=B_{\rm n,\Delta}$). $B_{\rm CIP}$, in this case, is less constrained, with a $1\sigma$ range  $-0.105<B_{\rm CIP}<0.183$.%This is different from the results using the DR10 data of \cite{Soumagnac2016}, where the authors obtained evidence at $3.2\sigma$ of $B_{\rm L,\Delta}>0.4$ (and evidence that $|B_{\rm L,\Delta}|>0.4$ at $3.7\sigma$) when allowing CIPs, which was an indication of the effect we search for.

 %in tension with the prediction of BL11. The $1\sigma$ upper limit
  %of % 21 bins $B_{\rm CIP}=$B_{\rm CIP}=6.76\times 10^{-3} and 2 sigma limit is $3.18\times 10^{-2}$,
 % $B_{\rm CIP}=2.0\times10^{-1}$ is less stringent by an order of magnitude compared to the unconstrained case. 

In the absence of CIPs, i.e. when $B_{\rm CIP}$ is set to zero,  $B_{\rm n,\Delta}$ is less constrained and the $1\sigma$ range $-6.0<B_{\rm n,\Delta}<4.0$ is consistent with both zero and the BL11 prediction for $B_{\rm L,\Delta}$. 

%The value of $\chi^2/dof$ for the joint best fit is high, $\chi^2/dof\approx2$. This could be due to an inconsistency of the model and the data at small angles, where the non-linearities may induce a systematic difference between the model and the data (see appendix~\ref{sec:scales}).

\begin{table*}
\begin{tabular}{|c|ccc|ccc|}
  &\multicolumn{3}{c|}{$B_{\rm CIP}\neq0$}&\multicolumn{3}{c|}{$B_{\rm CIP}=0$}\\ \hline 
  Parameter & med. & max. &$68.2\%$-range&med.&max.&$68.2\%$-range\\ \hline
  \rule[-2.ex]{0pt}{5ex} $B_{\rm n,t}$&$2.08$&$2.08$&$[2.03,2.12]$&$2.08$&$2.12$&$[2.02,2.13]$\\
  \rule[-2.ex]{0pt}{5ex} $B_{\rm L,t}$&$2.12$ &$2.10$&$[2.06,2.17]$&$2.13$&$2.18$&$[2.07,2.19]$\\%\cline{1-5}
  \rule[-2.ex]{0pt} {5ex}$B_{\rm L,\Delta}$&$-0.15$&$-2.66$&$[-0.91,0.74]$&$-1.00$&$-0.83$&$[-5.79,4.58]$\\ % \cline{1-5}
  \rule[-2.ex]{0pt}{5ex} $B_{\rm sys,L}\times10^{3}$& $-0.24$ &$-3.57$&$[-2.02,0.91]$&$0.85$&$0.84$&$[0.65,1.05]$\\
   \rule[-2.ex]{0pt}{5ex} $k_*$&$3.98$&$3.56$&$[3.26,4.71]$&$3.84$&$4.03$&$[2.34,5.01]$\\
  \rule[-2.ex]{0pt}{5ex} $A_{\rm MC}$&$4.08$&$4.48$&$[2.04,5.34]$&$3.78$&$5.75$&$[1.68,5.34]$\\
  \rule[-2.ex]{0pt}{5ex} $B_{\rm CIP}\times10^{2}$&$1.26$&$0.27$&$[-0.06,3.57]$&$-$&$-$&$-$\\
  \rule[-2.ex]{0pt}{5ex} $B_{\rm n,\Delta}$&$0.0484$&$-1.87$&$[-0.37,0.47]$&$-1.00$&$-0.89$&$[-5.79,4.59]$\\
     \rule[-2.ex]{0pt}{5ex} $B_{\rm sys,n}\times10^{4}$&$5.10$&$-15.0$&$[0.31,8.10]$&$6.80$&$6.72$&$[4.97,8.50]$\\
\end{tabular}
\caption{\label{table:paraXinXil_unconstrained}
{\bf Results of the model fitting, when using the CMASS sample and a model with unconstrained $B_{\rm L,\Delta}$} (i.e. the same model as in Soumagnac et al. 2016), for two configurations of the models: \{$B_{\rm CIP}\neq0, B_{\rm L,\Delta}\neq0$\} and \{$B_{\rm CIP}=0, B_{\rm L,\Delta}\neq0$\}. The table shows the maximum likelihood value, the median, and $68.2\%$ confidence range for each parameter, computed using the marginalised posterior distributions obtained with the Multinest code.}
%\end{ruledtabular}
\end{table*}
%\end{comment}

\begin{table*}
\begin{tabular}{|c|ccc|ccc|}
  &\multicolumn{3}{c|}{$B_{\rm CIP}\neq0$}&\multicolumn{3}{c|}{$B_{\rm CIP}=0$}\\ \hline 
  Parameter & med. & max. &$68.2\%$-range&med.&max.&$68.2\%$-range\\ \hline
\begin{comment} 
\rule[-2.ex]{0pt}{5ex}$B_{\rm n,t}$&$2.07$&$2.09$&$[2.01,2.12]$&$2.08$&$2.13$&$[2.02,2.13]$\\
\rule[-2.ex]{0pt}{5ex}$B_{\rm L,t}$&$2.12$&$2.14$&$[2.06,2.18]$&$2.13$&$2.19$&$[2.08,2.19]$\\
\rule[-2.ex]{0pt}{5ex}$B_{\rm sys,L}\times10^{3}$&$0.327$&$-1.79$&$[-1.11,0.989]$&$0.845$&$0.851$&$[0.664,1.04]$\\
\rule[-2.ex]{0pt}{5ex}$k_*$&$3.93$&$3.92$&$[3.14,4.71]$&$3.74$&$3.84$&$[2.2,4.96]$\\
\rule[-2.ex]{0pt}{5ex}$A_{\rm MC}$&$3.78$&$5.38$&$[1.74,5.28]$&$3.78$&$5.88$&$[1.74,5.34]$\\
\rule[-2.ex]{0pt}{5ex}$B_{\rm CIP}\times10^{2}$&$7.81$&$4.33$&$[-5.98,20.4]$&$-$&$-$&$-$\\
\rule[-2.ex]{0pt}{5ex}$B_{\rm n,\Delta}$&$0.00613$&$-0.47$&$[-0.201,0.213]$&$-1.4$&$-1.86$&$[-6.0,4.0]$\\
\rule[-2.ex]{0pt}{5ex}$B_{\rm sys,n}\times10^{4}$&$2.39$&$-15$&$[-9.3,8.09]$&$6.86$&$6.8$&$[4.99,8.47]$\\
\end{comment}

\rule[-2.ex]{0pt}{5ex}$B_{\rm n,t}$&$2.07$&$2.09$&$[2.02,2.12]$&$2.08$&$2.12$&$[2.02,2.14]$\\ 
 \rule[-2.ex]{0pt}{5ex}$B_{\rm L,t}$&$2.13$&$2.15$&$[2.07,2.18]$&$2.13$&$2.18$&$2.07,2.19]$\\ 
\rule[-2.ex]{0pt}{5ex}$B_{\rm sys,L}\times10^{3}$&$0.698$&$-1.33$&$[-0.251,1.05]$&$0.845$&$0.846$&$[0.661,1.04]$\\
\rule[-2.ex]{0pt}{5ex}$k_*$&$3.93$&$3.91$&$[3.16,4.7]$&$3.67$&$3.76$&$[2.2,4.97]$\\ 
\rule[-2.ex]{0pt}{5ex}$A_{\rm MC}$&$3.9$&$5.36$&$[1.86,5.28]$&$3.72$&$5.47$&$[1.68,5.34]$\\ 
\rule[-2.ex]{0pt}{5ex}$B_{\rm CIP}$&$0.048$&$0.0743$&$[-0.105,0.183]$&$-$&$-$&$-$\\ 
\rule[-2.ex]{0pt}{5ex}$B_{\rm n,\Delta}$&$-0.00765$&$-0.358$&$[-0.183,0.168]$&$-1.40$&$-1.42$&$[-6.2,4.2]$\\ 
\rule[-2.ex]{0pt}{5ex}$B_{\rm sys,n}\times10^{4}$&$5.39$&$-15$&$[-4.48,8.97]$&$6.83$&$6.77$&$[4.97,8.56]$\\ \hline

\end{tabular}
\caption{\label{table:paraXinXil_constrained}
{\bf Results of the model fitting, when using the CMASS sample and a model with $B_{\rm L,\Delta}=B_{\rm n,\Delta}$}, for two configurations of the models: \{$B_{\rm CIP}\neq0, B_{\rm L,\Delta}\neq0$\} and \{$B_{\rm CIP}=0, B_{\rm L,\Delta}\neq0$\}. The table shows the maximum likelihood value, the median, and $68.2\%$ confidence range for each parameter, computed using the marginalised posterior distributions obtained with the Multinest code.}
%\end{ruledtabular}
\end{table*}

\subsection{Model selection}\label{sec:sel}
To determine whether we detect a scale-dependent bias of the
luminosity correlation function requires answering the following
question: do the data support the inclusion of a non-zero
parameter $B_{\rm L,\Delta}$? Rather than a question of parameter
estimation (i.e. the determination of the most probable values for the extra parameters within the context of a single model), this is a question of model comparison between two models $\mathcal{M}$, with or without $B_{\rm
  L,\Delta}$. The parameter space $\Theta=\{B_{\rm n,t},B_{\rm
  L,t},B_{\rm n,\Delta},B_{\rm CIP},B_{\rm
  sys,n},B_{\rm sys,L},A_{\rm MC},k_*\}\equiv\{\phi,\psi\}$ is partitioned into the common parameters, $\phi=\{B_{\rm n,t},B_{\rm
  L,t},B_{\rm CIP},B_{\rm
  sys,n},B_{\rm sys,L},A_{\rm MC},k_*\}$, and the extra parameter $\psi=\{B_{\rm n,\Delta}\}$, describing the scale-dependent bias. The two models are nested, as defined in \citet{Verde2013}.

%\subsection{Model fitting in the presence of CIPs,  $B_{\rm CIP}\neq0$}
%The parameter space $ \Theta=\{B_{\rm n,t},B_{\rm L,t},B_{\rm L,\Delta},B_{\rm sys,L},B_{\rm sys,n},B_{\rm CIP},A_{\rm MC},k_*, B_{\rm n,\Delta}\}\equiv\{\phi,\psi\}$ is partitioned into the common parameters, $\phi=\{B_{\rm n,t},B_{\rm L,t},B_{\rm sys,L},B_{\rm sys,n},B_{\rm CIP},A_{\rm MC},k_*, B_{\rm n,\Delta}\}$, and the extra parameters, $\psi=\{B_{\rm L,\Delta}\}$, describing the scale-dependent bias.
Within a Bayesian framework \cite{Verde2013}, the key quantity for
comparing them is the Evidence (or model-averaged likelihood), $E=\int
Pr(\theta|\mathcal{M})Pr(\mathcal{D}|\theta,\mathcal{M})d\theta$. 

Our aim is to confront our degree of belief in the two different models $\mathcal{M}_{B_{\rm L,\Delta}\neq0}$ and $\mathcal{M}_{B_{\rm L,\Delta}=0}$ in the light of the data, i.e. to compare $Pr(\mathcal{M}_{B_{\rm L,\Delta}\neq0}| \mathcal{D})$ and $Pr(\mathcal{M}_{B_{\rm L,\Delta}=0}| \mathcal{D})$. 
Developing each term with Bayes' theorem $Pr(\mathcal{M}|\mathcal{D})=Pr(\mathcal{D}|\mathcal{M})\cdot Pr(\mathcal{M})/Pr(\mathcal{D})$, we can write
\begin{dmath}
\frac{Pr(\mathcal{M}_{B_{\rm L,\Delta}\neq0}|\mathcal{D})}{Pr(\mathcal{M}_{B_{\rm L,\Delta}=0}|\mathcal{D})}=\frac{Pr(\mathcal{D}|\mathcal{M}_{B_{\rm L,\Delta}\neq0})\cdot Pr(\mathcal{M}_{B_{\rm L,\Delta}\neq0})}{Pr(\mathcal{D})}\cdot \frac{Pr(\mathcal{D})}{Pr(\mathcal{D}|\mathcal{M}_{B_{\rm L,\Delta}=0})\cdot Pr(\mathcal{M}_{B_{\rm L,\Delta}=0})}\;.
\end{dmath}
Since we do not have any prior preference toward one of the models, $\frac{Pr(\mathcal{M}_{B_{\rm L,\Delta}\neq0})}{Pr(\mathcal{M}_{B_{\rm L,\Delta}=0})}$ is typically set to $1$ and the above ratio simplifies to 
\begin{dmath}
\frac{Pr(\mathcal{M}_{B_{\rm L,\Delta}\neq0}|\mathcal{D})}{Pr(\mathcal{M}_{B_{\rm L,\Delta}=0}|\mathcal{D})}=\frac{E_{B_{\rm L,\Delta}\neq0}}{E_{B_{\rm L,\Delta}=0}}\;.
\end{dmath}

%I band 
%b4 non zero
%ln(Z):              -49.574694
% ln(ev)=  -49.4443008197753      +/-  0.228523177144818
% Total Likelihood Evaluations:        34502
%b4 zero
%ln(Z):              -49.282339
%ln(ev)=  -49.1127551944057      +/-  0.227830457972594
%Total Likelihood Evaluations:        28588
%ln(E_{b_4\neq0}/E_{b_4=0})=-0.161961197753$
%G band 
%b4 non zero
%ln(Z):              -37.070421
% ln(ev)=  -36.9064677237549      +/-  0.205875368703404
% Total Likelihood Evaluations:        22903
%b4 zero
%ln(Z):              -37.235422
%ln(ev)=  -37.0285597841035      +/-  0.206409519822959
%Total Likelihood Evaluations:        21721

The ratio of the evidences can be calculated using the multimodal nested sampling algorithm, {\it MultiNest} \citep{Feroz2008} and are all shown in table~\ref{table:evidence} for the CMASS sample.% The same table is shown in the appendix for the LOWZ sample. 
 We use the slightly modified Jeffreys' scale \citep{Jeffrey1961, Kass1995,Verde2013} shown in the appendix in table~\ref{table:Jef}, which classifies Evidence ratios from ''not worth a bare mention'' to ``highly significant'', to interpret these values. 

In the unconstrained case, i.e. within the model used by \cite{Soumagnac2016}, the results are as follows. (1) When including CIPs in the model, the evidence ratio $\ln(E_{B_{\rm L,\Delta}\neq0}/E_{B_{\rm L,\Delta}=0})=-1.04\pm0.21$ corresponds to no evidence toward a non-zero $B_{\rm L,\Delta}$ over a zero effect. (2) In the absence of CIPs (i.e., when setting $B_{\rm CIP} = 0$), the data strongly privilege $B_{\rm L,\Delta}\neq0$, i.e. the presence of the effect we search for. This result is the opposite of the DR10 result presented in \cite{Soumagnac2016}, where a strong evidence toward a non-zero $B_{\rm L,\Delta}\neq0$ was measured when $B_{\rm CIP}\neq0$ and disappeared in the absence of CIPs. 

In the constrained case, i.e. when adding the more realistic constraint $B_{\rm L,\Delta}=B_{\rm n,\Delta}$ to the model by \cite{Soumagnac2016}, the calculation of the evidence ratio leads to different conclusions. (1) In the presence of CIPs, the data privilege a zero $B_{\rm L,\Delta}$. The evidence ration $\ln(E_{B_{\rm L,\Delta}\neq0}/E_{B_{\rm L,\Delta}=0})=-2.56\pm0.20$ corresponds - according to the Jeffrey's table~\ref{table:Jef} shown in the appendix- to ``substantial'' evidence for $B_{\rm L,\Delta}=0$. (2) In the absence of CIPs, the evidence for $B_{\rm L,\Delta}=0$ goes away: like in the DR10 analysis, the data do not privilege the effect we search for over the $B_{\rm L,\Delta}=0$ case.
 
\begin{table}
\begin{tabular}{|c||c|c|}\hline

  \rule[-2.ex]{0pt}{5ex}  &Unconstrained case& Constrained case\\
 \rule[-2.ex]{0pt}{5ex}  && ($B_{\rm L,\Delta}=B_{\rm n,\Delta}$)\\ \hline
 \rule[-2.ex]{0pt}{5ex}  $B_{\rm CIP}\neq 0$&$-57.88\pm0.11$&$-57.08\pm0.10$\\
 \rule[-2.ex]{0pt}{5ex}  $B_{\rm L,\Delta}\neq 0$&&\\ \hline
  \rule[-2.ex]{0pt}{5ex}  $B_{\rm CIP}\neq 0$&$-56.84\pm 0.10$&$-54.52\pm0.10$\\
  \rule[-2.ex]{0pt}{5ex} $B_{\rm L,\Delta}=0$&&\\ \hline
  \rule[-2.ex]{0pt}{5ex}  $B_{\rm CIP}=0$&$ -56.18\pm0.10$&$-54.83\pm0.10$\\
  \rule[-2.ex]{0pt}{5ex} $B_{\rm L,\Delta}\neq 0$&&\\ \hline
  \rule[-2.ex]{0pt}{5ex}  $B_{\rm CIP}=0$&$-65.95\pm0.11$&$-54.75\pm0.10$\\
 \rule[-2.ex]{0pt}{5ex}  $B_{\rm L,\Delta}=0$&&\\ \hline
 
 \end{tabular}
 \caption{Logarithm of the Bayesian evidence calculated with the multimodal nested sampling algorithm, {\it MultiNest} \citep{Feroz2008}, using the CMASS sample. \label{table:evidence}}
%\end{ruledtabular}
\end{table}

\section{Conclusion}\label{sec:cclrem}

We have compared the large scale distribution of mass and light, through measurement of the number-weighted and luminosity-weighted correlation functions $\xi_{\rm n}$ and $\xi_{\rm L}$, in the CMASS sample of DR12, the latest and last public data release from the SDSS-III Baryon Oscillation Spectroscopic Survey (BOSS). We have detailed the method for the detection, with a data set containing 3-D positions and photometry, of the modulation of the large scales ratio of baryonic matter to total matter ($\delta_{b}/\delta_{\rm tot}$), from BAOs. Within the framework of a model presented in \citet{Barkana2011} (BL11), which we have reformulated and specified, this modulation is characterised by a parameter, $B_{\rm L,\Delta}$, which we have measured in the BOSS CMASS DR12 data. 

When compensated isocurvature perturbations (CIPs) are included in the model, the DR12 result is in tension with the strong evidence toward the effect we search for, which was observed in DR10. Indeed, the DR12 data is consistent either with a null detection - when using the same model as in \citealt{Soumagnac2016} - or slightly privilege a model with no effect - when adding to the model an additional constraint that reflects our knowledge of the survey flux limit. If we do not include CIPs in the models and use our knowledge of the survey flux limit, we obtain a null detection of the effect, consistent with both $B_{\rm L,\Delta}=0$, and the theoretical $B_{\rm L,\Delta}$ predicted in previous theoretical studies. 

%When using the same model as in \cite{Soumagnac2016}, the DR12 and DR10 results are in tension, as the DR12 data seem to be consistent either with a zero effect or a null detection, depending on whether compensated isocurvature perturbations (CIPs) are included in the model or not. However, this result changes when we add to the model an additional constraint that reflects our knowledge of the survey flux limit. Then, depending on whether we include CIPs in the model, we either obtain substantial evidence for no effect or a null detection consistent with both $B_{\rm n,\Delta}=0$, and the theoretical $B_{\rm n,\Delta}$ predicted in previous theoretical studies. 
%
%Future observational efforts, such as DESI, the Dark Energy Spectroscopic Instrument \citep{Levi2013}, will provide more accurate data. We expect new data sets to prove or disprove the predicted effect if our error bars on $b_4$ decrease by a factor of 5. A better account for the non-linear effects, e.g. with reconstruction, could improve the goodness of the fit and increase the evidence for a non-zero $b_4$. Future developments of this work should also include a deeper study of the feasibility of the detection, with current and future data sets, e.g. with simulations.  

As noted in \cite{Soumagnac2016}, disentangling the various effects at stake is
difficult. On the one hand, the model of equations~\ref{eq:xin_new}
and~\ref{eq:xil} shows that any ability to set a limit on CIPs depends
on a definitive detection of non-zero $B_{\rm L,\Delta}$ (and/or $B_{\rm n,\Delta}$). Conversely, the presence of a significant CIP
term in the fit affects the range of $B_{\rm L,\Delta}$ and $B_{\rm n,\Delta}$ values. Trying to measure two novel
effects (one of them expected but with an uncertain amplitude, the
other highly speculative) when they are entangled is tricky. Another difficulty comes from the fact that $\hat{\xi}_{\rm
  CIP}$ has a smooth shape (in contrast with BAO-scale features in
$\xi_{\rm tot}$ and $\xi_{\rm add}$), and such a slowly-varying term
may more easily be emulated by systematic effects; we note that
standard BAO measurements (e.g., \citep{Percival2014}) typically add
several such ``nuisance'' terms, which are necessary to get good fits
to the data, do not significantly affect the BAO peak/trough
positions, but are not theoretically well-understood. 

We believe that both the DR10 and DR12 results demonstrate that current data
are on the threshold of detecting the BAO-induced modulation and
setting strong limits on
CIPs. %This lack of evidence is reflected by the evidence ratio we measure in section~\ref{sec:sel}.
In particular, future observational efforts, such as the Dark Energy Spectroscopic
Instrument (DESI) \citep{Levi2013}, will provide more accurate data. The large number of galaxies will reduce the
  statistical error on the correlation function measurement and
  increase the redshift coverage. The better quality imaging will
  reduce the error on the luminosity measurement and subsequently on
  $\xi_{\rm L}$. More robust theoretical modeling as well as new data sets may allow to definitively verify or rule out
the predicted effect.

 %A better account for the non-linear effects, e.g. with reconstruction, could improve the goodness of the fit and increase the evidence for a non-zero $B_{\rm L,\Delta}$. Future developments of this work should also include a deeper study of the feasibility of the detection, with current and future data sets, e.g. with simulations.

\section{Acknowledgments}

M.T.S. acknowledges support by a grant from IMOS/ISA, the Ilan Ramon fellowship from the Israel Ministry of Science and Technology and the Benoziyo center for Astrophysics at the Weizmann Institute of Science.

CGS acknowledges support from the National Research Foundation of Korea (NRF-2017R1D1A1B03034900). 

Funding for SDSS-III has been provided by the Alfred P. Sloan Foundation, the Participating Institutions, the National Science Foundation, and the U.S. Department of Energy Office of Science. The SDSS-III web site is http://www.sdss3.org/. SDSS-III is managed by the Astrophysical Research Consortium for the Participating Institutions of the SDSS-III Collaboration including the University of Arizona, the Brazilian Participation Group, Brookhaven National Laboratory, Carnegie Mellon University, University of Florida, the French Participation Group, the German Participation Group, Harvard University, the Instituto de Astrofisica de Canarias, the Michigan State/Notre Dame/JINA Participation Group, Johns Hopkins University, Lawrence Berkeley National Laboratory, Max Planck Institute for Astrophysics, Max Planck Institute for Extraterrestrial Physics, New Mexico State University, New York University, Ohio State University, Pennsylvania State University, University of Portsmouth, Princeton University, the Spanish Participation Group, University of Tokyo, University of Utah, Vanderbilt University, University of Virginia, University of Washington, and Yale University.

%
%Funding for SDSS-III has been provided by the Alfred P. Sloan Foundation, the Participating Institutions, the National Science Foundation, and the U.S. Department of Energy Office of Science. The SDSS-III web site is http://www.sdss3.org/. SDSS-III is managed by the Astrophysical Research Consortium for the Participating Institutions of the SDSS-III Collaboration including the University of Arizona, the Brazilian Participation Group, Brookhaven National Laboratory, Carnegie Mellon University, University of Florida, the French Participation Group, the German Participation Group, Harvard University, the Instituto de Astrofisica de Canarias, the Michigan State/Notre Dame/JINA Participation Group, Johns Hopkins University, Lawrence Berkeley National Laboratory, Max Planck Institute for Astrophysics, Max Planck Institute for Extraterrestrial Physics, New Mexico State University, New York University, Ohio State University, Pennsylvania State University, University of Portsmouth, Princeton University, the Spanish Participation Group, University of Tokyo, University of Utah, Vanderbilt University, University of Virginia, University of Washington, and Yale University.
\footnotesize{
  \bibliographystyle{mn2e}
\providecommand{\eprint}[1]{\href{http://arxiv.org/abs/#1}{arXiv:#1}}	
  \bibliography{thesis}
}

\appendix

\section[]{Interpretation of the Evidence ratio $ln(E_{B_{\rm L,\Delta}\neq0}/E_{B_{\rm L,\Delta}=0})$}

In table~\ref{table:Jef}, we show the slightly modified Jeffreys' scale \citep{Jeffrey1961, Kass1995,Verde2013} which we use to interpret the evidence shown is table~\ref{table:evidence}.

   \begin{table}
\centering
    \begin{tabular}{| c || c|c|  }
       \hline
  $ln(E_{B_{\rm L,\Delta}\neq0}/E_{B_{\rm L,\Delta}=0})$ & interpretation & betting odds\\ \hline
 \rule[-2.ex]{0pt}{5ex}  $<1$ & not worth a bare mention & $<3:1$\\ \hline
 \rule[-2.ex]{0pt}{5ex}  $1 - 2.5$ & substancial & $~3:1$\\ \hline
  \rule[-2.ex]{0pt}{5ex}  $2.5 - 5$ &strong & $>12:1$\\ \hline
   \rule[-2.ex]{0pt}{5ex}  $>5$ & highly significant & $>150:1$\\ \hline
     \end{tabular}
 \caption{The slightly modified Jeffrey's scale we use to interpret the Evidence ratio $ln(E_{B_{\rm L,\Delta}\neq0}/E_{B_{\rm L,\Delta}=0})$.}     
\label{table:Jef}
     \end{table}

\bsp
\label{lastpage}
\end{document}